\documentclass[twocolumn]{aastex62}

\usepackage{graphicx}

\usepackage{multirow}
\usepackage{url}
\usepackage{txfonts}

\usepackage{microtype}

\shorttitle{The VLA Sky Survey}
\shortauthors{VLA Sky Survey Science Group}

\begin{document}


\title{The Karl G. Jansky Very Large Array Sky Survey (VLASS): science case and survey design.}



\author[0000-0002-3032-1783]{M. Lacy}  
\affiliation{National Radio Astronomy Observatory, 520 Edgemont Road, Charlottesville, VA 22903, USA} 
\author{S. A. Baum} 
\affiliation{Department of Physics and Astronomy, University of Manitoba, Winnipeg, MB R3T 2N2, Canada}
\author[0000-0002-7570-5596]{C. J. Chandler} 
\affiliation{National Radio Astronomy Observatory, 1003 Lopezville Road, Socorro, NM 87801, USA} 
\author[0000-0002-2878-1502]{S. Chatterjee} 
\affiliation{Cornell Center for Astrophysics and Planetary Science and Department of Astronomy, Cornell University, Ithaca, NY 14853, USA} 
\author{T. E. Clarke} 
\affiliation{Naval Research Laboratory Remote Sensing Division, Code 7213, 4555 Overlook Avenue SW, Washington, DC 20375, USA} 
\author[0000-0003-2823-360X]{S. Deustua} 
\affiliation{Space Telescope Science Institute, 3700 San Martin Dr, Baltimore, MD 21218, USA} 
\author{J. English} 
\affiliation{Department of Physics and Astronomy, University of Manitoba, Winnipeg, MB R3T 2N2, Canada}
\author{J. Farnes} 
\affiliation{Oxford e-Research Centre, Keble Road, Oxford OX1 3QG, UK}
\author{B. M. Gaensler} 
\affiliation{Dunlap Institute for Astronomy and Astrophysics, University of Toronto, 50 St George Street, Toronto, ON M5S 3H4, Canada} 
\author{N. Gugliucci} 
\affiliation{Department of Physics, St Anselm College, 100 Saint Anselm Drive, Manchester, NH  03102, USA}
\author[0000-0002-7083-4049]{G. Hallinan} 
\affiliation{California Institute of Technology, 1200 East California Boulevard, Pasadena, CA 91125, USA} 
\author[0000-0002-8990-1811]{B. R. Kent} 
\affiliation{National Radio Astronomy Observatory, 520 Edgemont Road, Charlottesville, VA 22903, USA} 
\author{A. Kimball} 
\affiliation{National Radio Astronomy Observatory, 1003 Lopezville Road, Socorro, NM 87801, USA} 
\author[0000-0002-4119-9963]{C. J. Law} 
\affiliation{California Institute of Technology, 1200 East California Boulevard, Pasadena, CA 91125, USA} 
\affiliation{Department of Astronomy and Radio Astronomy Lab, University of California, Berkeley, CA 94720, USA} 
\author{T. J. W. Lazio} 
\affiliation{Jet Propulsion Laboratory, California Institute of Technology, Pasadena, CA 91109, USA} 
\author{J. Marvil} 
\affiliation{National Radio Astronomy Observatory, 1003 Lopezville Road, Socorro, NM 87801, USA} 
\author[0000-0001-8906-7866]{S. A. Mao} 
\affiliation{Max-Planck-Institut f\"ur Radioastronomie, Auf dem H\"ugel 69, 53121 Bonn, Germany} 
\author{D. Medlin}
\affiliation{National Radio Astronomy Observatory, 1003 Lopezville Road, Socorro, NM 87801, USA}
\author[0000-0002-2557-5180]{K. Mooley} 
\affiliation{California Institute of Technology, 1200 East California Boulevard, Pasadena, CA 91125, USA} 
\author{E. J. Murphy} 
\affiliation{National Radio Astronomy Observatory, 520 Edgemont Road, Charlottesville, VA 22903, USA} 
\author{S. Myers} 
\affiliation{National Radio Astronomy Observatory, 1003 Lopezville Road, Socorro, NM 87801, USA} 
\author{R. Osten} 
\affiliation{Space Telescope Science Institute, 3700 San Martin Dr, Baltimore, MD 21218, USA} 
\author{G.T. Richards} 
\affiliation{Department of Physics, Drexel University, 3141 Chestnut Street, Philadelphia, PA 19104, USA} 
\author{E. Rosolowsky} 
\affiliation{Department of Physics, University of Alberta, CCIS 4-181, Edmonton AB T6G 2E1, Canada} 
\author{L. Rudnick} 
\affiliation{Minnesota Institute for Astrophysics, School of Physics and Astronomy, University of Minnesota, 116 Church Street SE, Minneapolis, MN 55455, USA} 
\author{F. Schinzel} 
\affiliation{National Radio Astronomy Observatory, 1003 Lopezville Road, Socorro, NM 87801, USA} 
\author[0000-0001-6682-916X]{G. R. Sivakoff} 
\affiliation{Department of Physics, University of Alberta, CCIS 4-181, Edmonton AB T6G 2E1, Canada} 
\author{L. O. Sjouwerman} 
\affiliation{National Radio Astronomy Observatory, 1003 Lopezville Road, Socorro, NM 87801, USA} 
\author{R. Taylor} 
\affiliation{Department of Astronomy, University of Cape Town, Private Bag X3, Rondebosch 7701, South Africa} 
\affiliation{Department of Physics and Astronomy, University of the Western Cape, Private Bag X17, Bellville 7535, South Africa}
\author{R. L. White} 
\affiliation{Space Telescope Science Institute, 3700 San Martin Dr, Baltimore, MD 21218, USA}
\author[0000-0001-9720-7398]{J. Wrobel} 
\affiliation{National Radio Astronomy Observatory, 1003 Lopezville Road, Socorro, NM 87801, USA} 
\author[0000-0003-4873-1681]{H. Andernach}
\affiliation{Departamento de Astronom\'{i}a, DCNE, Universidad de Guanajuato, Apdo. Postal 144, 36000 Guanajuato, Guanajuato, Mexico}
\author{A. J. Beasley} 
\affiliation{National Radio Astronomy Observatory, 520 Edgemont Road, Charlottesville, VA 22903, USA} \author{E. Berger} 
\affiliation{Harvard Astronomy Department, 60 Garden Street, MS 46, Cambridge, MA 02138, USA} 
\author{S. Bhatnager} 
\affiliation{National Radio Astronomy Observatory, 1003 Lopezville Road, Socorro, NM 87801, USA} 
\author{M. Birkinshaw}
\affiliation{HH Wills Physics Laboratory, University of Bristol, Tyndall Avenue, Bristol BS8 1TL, UK}
\author[0000-0003-4056-9982]{G.C. Bower} 
\affiliation{Academia Sinica Institute of Astronomy and Astrophysics, 645 N. A$'$ohoku Place, Hilo, HI 96720, USA} 
\author[0000-0002-0167-2453]{W. N. Brandt} 
\affiliation{Department of Astronomy and Astrophysics, 525 Davey Lab, The Pennsylvania State University, University Park, PA 16802, USA} 
\affiliation{Institute for Gravitation and the Cosmos, The Pennsylvania State University, University Park, PA 16802, USA}
\affiliation{Department of Physics, 104 Davey Lab, The Pennsylvania State University, University Park, PA 16802, USA}
\author{S. Brown} 
\affiliation{The University of Iowa, Department of Physics and Astronomy, 203 Van Allen Hall, Iowa City, IA 52242, USA} 
\author{S. Burke-Spolaor} 
\affiliation{Department of Physics and Astronomy, West Virginia University, White Hall, Morgantown, WV 26506, USA} 
\affiliation{Center for Gravitational Waves and Cosmology, West Virginia University, Morgantown, WV, USA.}
\author{B. J. Butler}
\affiliation{National Radio Astronomy Observatory, 1003 Lopezville Road, Socorro, NM 87801, USA}
\author{J. Comerford} 
\affiliation{Department of Astrophysical and Planetary Sciences, University of Colorado, Boulder, CO 80309, USA} 
\author{P. B. Demorest}
\affiliation{National Radio Astronomy Observatory, 1003 Lopezville Road, Socorro, NM 87801, USA}
\author{H. Fu} 
\affiliation{The University of Iowa, Department of Physics and Astronomy, 203 Van Allen Hall, Iowa City, IA 52242, USA} 
\author{S. Giacintucci} 
\affiliation{Naval Research Laboratory Remote Sensing Division, Code 7213, 4555 Overlook Avenue SW, Washington, DC 20375, USA} 
\author{K. Golap}
\affiliation{National Radio Astronomy Observatory, 1003 Lopezville Road, Socorro, NM 87801, USA}
\author{T. G\"uth}
\affiliation{National Radio Astronomy Observatory, 1003 Lopezville Road, Socorro, NM 87801, USA}
\author[0000-0002-3733-2565]{C. A. Hales}
\altaffiliation{Marie Sk\l{}odowska-Curie Fellow}
\affiliation{School of Mathematics, Statistics and Physics, Newcastle University, Newcastle upon Tyne NE1 7RU, UK}
\affiliation{National Radio Astronomy Observatory, 1003 Lopezville Road, Socorro, NM 87801, USA}
\author{R. Hiriart}
\affiliation{National Radio Astronomy Observatory, 1003 Lopezville Road, Socorro, NM 87801, USA}
\author{J. Hodge} 
\affiliation{Leiden Observatory, Leiden University, PO Box 9513, NL-2300 RA Leiden, the Netherlands} 
\author{A. Horesh} 
\affiliation{Racah Institute of Physics, The Hebrew University of Jerusalem, Jerusalem, 91904, Israel} 
\author{\v{Z}.~Ivezi\'{c}}
\affiliation{University of Washington, Dept. of Astronomy, Box 351580, Seattle, WA 98195, USA}
\author{M. J. Jarvis} 
\affiliation{Astrophysics, University of Oxford, Denys Wilkinson Building, Keble Road, Oxford OX1 3RH, UK}
\affiliation{Department of Physics and Astronomy, University of the Western Cape, Private Bag X17, Bellville 7535, South Africa}
\author{A. Kamble} 
\affiliation{Center for Astrophysics $\mid$ Harvard and Smithsonian, 60 Garden Street, Cambridge, MA 02138, USA}
\author{N. Kassim} 
\affiliation{Naval Research Laboratory Remote Sensing Division, Code 7213, 4555 Overlook Avenue SW, Washington, DC 20375, USA} 
\author[0000-0003-0049-5210]{X. Liu} 
\affiliation{Department of Astronomy, University of Illinois at Urbana-Champaign, Urbana, IL 61801, USA}
\author[0000-0002-5635-3345]{L. Loinard}
\affiliation{Instituto de Radioastronom\'{\i}a y Astrof\'{\i}sica, Universidad Nacional Aut\'onoma de M\'exico, 58089, Michoac\'an, Mexico}
\affiliation{Instituto de Astronomía, Universidad Nacional Aut\'onoma de M\'exico, 04510 Ciudad de M\'exico, Mexico}
\author[0000-0003-2899-2374]{D. K. Lyons}
\affiliation{National Radio Astronomy Observatory, 1003 Lopezville Road, Socorro, NM 87801, USA}
\author{J. Masters} 
\affiliation{National Radio Astronomy Observatory, 520 Edgemont Road, Charlottesville, VA 22903, USA} 
\author{M. Mezcua}
\affiliation{Institute of Space Sciences (ICE, CSIC), Campus UAB, Carrer de Magrans, 08193 Barcelona, Spain}
\affiliation{Institut d'Estudis Espacials de Catalunya (IEEC), Carrer Gran Capità, 08034 Barcelona, Spain}
\author{G. A. Moellenbrock}
\affiliation{National Radio Astronomy Observatory, 1003 Lopezville Road, Socorro, NM 87801, USA}
\author[0000-0003-3816-5372]{T. Mroczkowski} 
\affiliation{ESO - European Southern Observatory, Karl-Schwarzschild-Str.\ 2, DE-85748 Garching b. M\"unchen, Germany} 
\author{K. Nyland} 
\affiliation{National Research Council, resident at the Naval Research Laboratory, Washington, DC 20375, USA}
\author[0000-0001-6421-054X]{C. P. O'Dea} 
\affiliation{Department of Physics and Astronomy, University of Manitoba, Winnipeg, MB R3T 2N2, Canada} 
\author[0000-0002-3968-3051]{S. P. O'Sullivan}
\affiliation{Hamburger Sternwarte, Universit\"{a}t Hamburg, Gojenbergsweg 112, D-21029 Hamburg, Germany.}
\author{W. M. Peters} 
\affiliation{Naval Research Laboratory Remote Sensing Division, Code 7213, 4555 Overlook Avenue SW, Washington, DC 20375, USA}
\author{K. Radford}
\affiliation{National Radio Astronomy Observatory, 1003 Lopezville Road, Socorro, NM 87801, USA}
\author{U. Rao}
\affiliation{National Radio Astronomy Observatory, 1003 Lopezville Road, Socorro, NM 87801, USA}
\author{J. Robnett} 
\affiliation{National Radio Astronomy Observatory, 1003 Lopezville Road, Socorro, NM 87801, USA}
\author{J. Salcido} 
\affiliation{National Radio Astronomy Observatory, 1003 Lopezville Road, Socorro, NM 87801, USA}
\author{Y. Shen} 
\affiliation{Department of Astronomy, University of Illinois at Urbana-Champaign, Urbana, IL 61801, USA}
\affiliation{National Center for Supercomputing Applications, University of Illinois at Urbana-Champaign, Urbana, IL 61801, USA}
\author{A. Sobotka} 
\affiliation{National Radio Astronomy Observatory, 1003 Lopezville Road, Socorro, NM 87801, USA} 
\author{S. Witz} 
\affiliation{National Radio Astronomy Observatory, 1003 Lopezville Road, Socorro, NM 87801, USA} 
\author[0000-0002-6748-0577]{M. Vaccari}
\affiliation{Department of Physics and Astronomy, University of the Western Cape, Private Bag X17, Bellville 7535, South Africa}
\affiliation{INAF - Instituto di Radioastronomia, via Gobetti 101, I-40129 Bologna, Italy}
\author{R. J. van Weeren}
\affiliation{Leiden Observatory, Leiden University, PO Box 9513, NL-2300 RA Leiden, the Netherlands}
\author{A. Vargas}
\affiliation{National Radio Astronomy Observatory, 1003 Lopezville Road, Socorro, NM 87801, USA}
\author[0000-0003-3734-3587]{P. K. G. Williams} 
\affiliation{Center for Astrophysics $\mid$ Harvard and Smithsonian, 60 Garden Street, Cambridge, MA 02138, USA}
\author[0000-0001-9163-0064]{I. Yoon}
\affiliation{National Radio Astronomy Observatory, 520 Edgemont Road, Charlottesville, VA 22903, USA}

\correspondingauthor{Mark Lacy}
\email{mlacy@nrao.edu}


\begin{abstract}
The Very Large Array Sky Survey (VLASS) is a synoptic, all-sky radio sky survey with a unique combination of high angular resolution ($\approx$2\farcs5), sensitivity (a 1$\sigma$ goal of 70\,$\muup$Jy/beam in the coadded data), full linear Stokes polarimetry, time domain coverage, and wide bandwidth (2--4~GHz). The first observations began in September 2017, and observing for the survey will finish in 2024. VLASS will use approximately 5500 hours of time on the Karl G.\ Jansky Very Large Array (VLA) to cover the whole sky visible to the \hbox{VLA} (Declination $>-40\degr$), a total of 33\,885~deg${}^2$. The data will be taken in three epochs to allow the discovery of variable and transient radio sources. The survey is designed to engage radio astronomy experts, multi-wavelength  astronomers, and citizen scientists alike. By utilizing an ``on the fly" interferometry mode, the observing overheads are much reduced compared to a conventional pointed survey. In this paper, we present the science case and observational strategy for the survey, and also results from early survey observations.

\end{abstract}





\section{Introduction}

\subsection{Motivation}

The advent of wide bandwidth backends has increased the sensitivity of radio interferometers such as the Karl G. Jansky Very Large Array \citep[VLA;][]{2011ApJ...739L...1P} to radio continuum emission by a factor of several, as well as providing instantaneous frequency-dependent flux density and polarization information across bandwidths covering factors of order two in frequency. When coupled with another important innovation, on the fly mosaicking \citep[OTFM, see the OTFM guide at https://go.nrao.edu/vlass-mosaicking and ][]{2018ApJ...857..143M,2019ApJ...870...25M}, the sky can be rapidly imaged to a relatively deep limit at high angular resolution, providing both polarization and spectral index information. Such advances justify another major survey with the VLA, complementing the earlier 
NRAO VLA Sky Survey \citep[NVSS;][]{nvss} and Faint Images of the Radio Sky at Twenty centimetres \citep[FIRST;][]{first95} surveys.

A new radio survey with the VLA was further motivated by major surveys now being undertaken in the optical and near-infrared. Large-format detectors in these regimes have greatly expanded 
the capabilities of surveys, making the combination of depth and frequent re-observation possible over tens of thousands of square degrees. These capabilities have allowed synoptic surveys that characterize stellar and galaxy populations in unprecedented detail. Optical photometric surveys already available include the Sloan Digital Sky Survey \citep[SDSS;][]{2018ApJS..235...42A}, HyperSuprimeCam \citep[HSC;][]{2018PASJ...70S...8A}, the Dark Energy Camera Legacy Survey (DECaLS) and related surveys \citep{2018arXiv180408657D}, the Dark Energy Survey \citep[DES;][]{2018arXiv180103181A} and the Panoramic Survey Telescope and Rapid Response System \citep[Pan-STARRS;][]{2016arXiv161205560C}. In the near-infrared, surveys such as the UKIRT Infrared Deep Sky Survey \citep[UKIDSS;][]{2007MNRAS.379.1599L} and surveys made with the Visible and Infrared Survey Telescope for Astronomy \citep[VISTA;][]{2015A&A...575A..25S} exist over much of the sky, and there is also an all-sky infrared survey by the {\em Wide-field Infrared Survey Explorer (WISE)} at 3.5, 4.6, 12 and 22$\, \muup$m \citep{2010AJ....140.1868W}. Spectroscopic surveys have also taken advantage of new technologies, including the Baryon Oscillation Spectroscopic Survey \citep[BOSS;][]{2013AJ....145...10D}. 

Following the conclusion of the Expanded VLA (EVLA) Project in~2013,
the National Radio Astronomy Observatory (NRAO) announced that it
would consider a new radio sky survey to exploit the new and upgraded capabilities of the
telescope. 

\begin{figure*}[tb]
\centering
\includegraphics[width=0.9\textwidth]{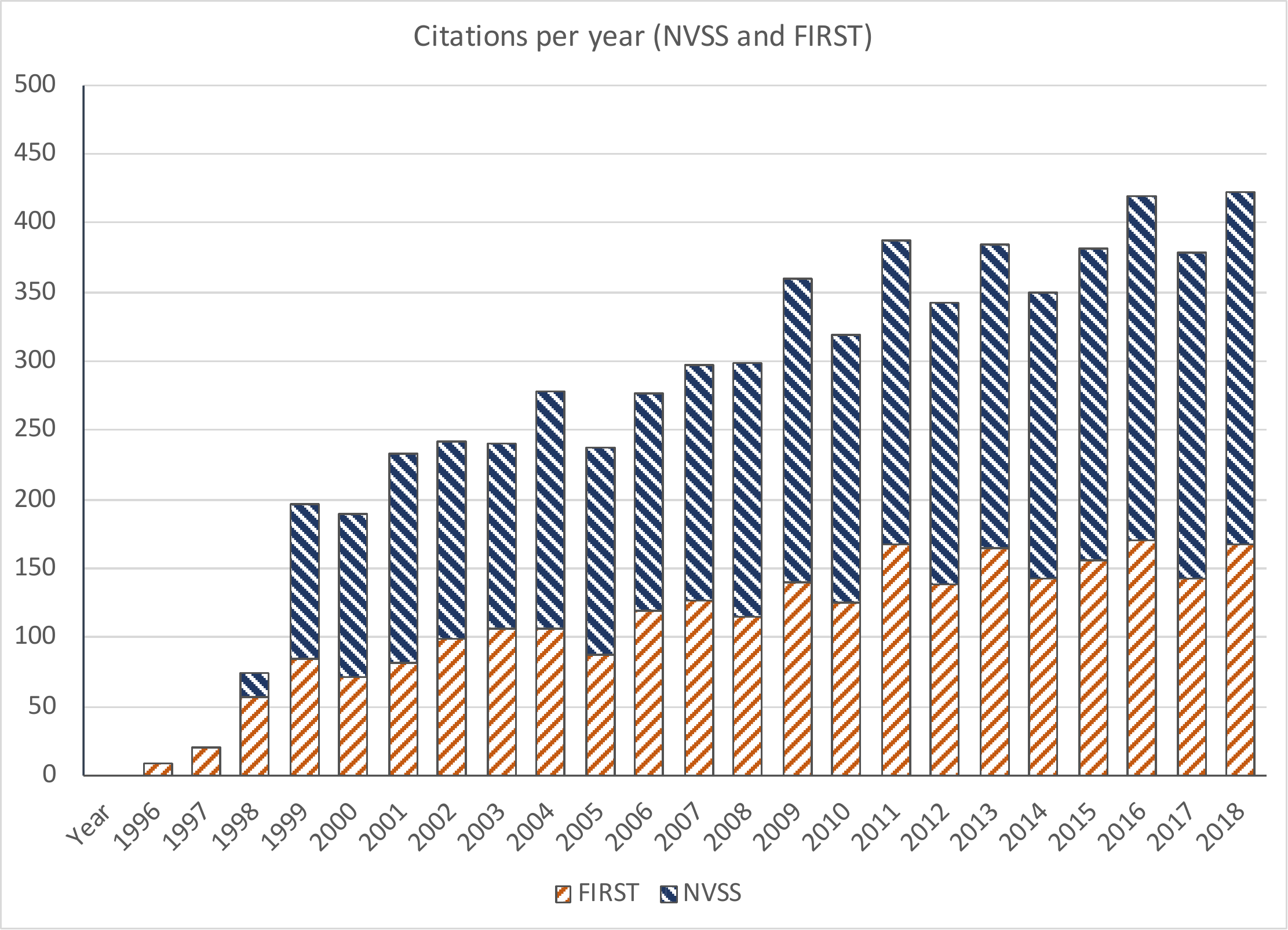}
\caption{Surveys can have considerable ``staying power'', as
illustrated by the citation statistics for two surveys conducted using the Very Large Array.  The NRAO VLA Sky Survey \protect\citep[\hbox{NVSS}, ][]{nvss}, and the Faint Images of the Radio Sky at Twenty
Centimeters \protect\citep[\hbox{FIRST}, ][]{first95,first97}.  Even though both
surveys are approaching their second decade since completion, the
number of citations, as indexed by the Astrophysics Data System shown here as a stacked histogram of NVSS and FIRST citations (which may include some double counting from papers that use both surveys),
is holding steady or increasing.}
\label{fig:metrics2}
\end{figure*}
Radio interferometry is a relatively specialized branch of astronomy, and having science-ready survey data products available is particularly valuable. This is illustrated by the very high usage of NVSS and FIRST data products. A quantitative measure comes from statistics gathered from the FIRST image server,\footnote{\url{http://third.ucllnl.org/cgi-bin/firstcutout}
}
which provides JPEG or FITS cutouts extracted from the FIRST survey at
user-specified positions.  In a recent 18~month interval, the server
delivered (on average) more than 7500 images per day, or one image every 12s
($>$95\% of these are scripted queries rather than individual users going to the website).  Each image served is equivalent to a three-minute VLA observation (the exposure time required
to reach the FIRST depth). Every 10~days, the effective
exposure time ($7500~\mathrm{images}~\mathrm{per\;day} \times
3~\mathrm{min.}~\mathrm{per\;image} \times 10~\mathrm{days}$) provided
is equal to the 4000~hr used originally to conduct the FIRST survey.
Figure \ref{fig:metrics2} shows 
that such surveys also provide a very high return in terms of citations.

\subsection{Survey Design}
The concept for VLASS was developed through a community-led  process, beginning with a public workshop at the January 2014 meeting of the American Astronomical Society, the submission of 22 white papers\footnote{https://go.nrao.edu/vlass-whitepapers} (on which much of Section \ref{sec:science} of this paper is based) and a competitive debate within a community-led Survey Science Group and its constituent working groups. An internal NRAO scientific and technical review, followed by a Community Review, provided important additional input. The design of VLASS paid extremely close attention to the Square Kilometre Array (SKA) pathfinders, leading to a survey that will both stand alone, yet also be complementary to those surveys.  

\begin{table*}[tbh]
\begin{center}
\caption{Summary of the VLASS design goals
\label{tab:all-sky}}
{\footnotesize
\begin{tabular}{ll}
\noalign{\hrule\hrule}
\textbf{Parameter} & \textbf{Value} \\
\noalign{\hrule\hrule}
 Total Area & 33885~deg$^{2}$, $\delta > -40^\circ$ \\
  Cadence   & 3 epochs, separated by $\sim$32~months \\
  Frequency Coverage & 2 -- 4 GHz\\          
  Angular Resolution & 2\farcs5 (B/BnA configurations) \\
  Continuum Image rms (Stokes~I) & $\sigma_I = 70\,\muup$Jy/beam combined \\
                                 & $\sigma_I = 120\,\muup$Jy/beam per-epoch \\
  On-Sky Integration Time & 4504~hr total \\
                   & 1501~hr per epoch, \\
  Scheduled Time (w/ 19\% overhead + & 5520~hr in total observing, \\
  ~~~~~~3\% for failed observations) & 1840~hr per epoch \\
  Estimated total detections$^{\dag}$ & $\sim$5,300,000\\
\noalign{\hrule\hrule}
\end{tabular}

\noindent
{\scriptsize $^{\dag}$Estimated number of individual source components (including parts of single resolved sources) above 5$\sigma \approx 350\,\muup{\rm Jy}$ in the final cumulative images.} 
}
\end{center}
\end{table*}

The VLASS design that emerged from the above process is an approximately 5500~hr survey covering the whole sky visible to the VLA at high angular resolution ($\approx 2.5$ arcsec) in three epochs, as summarized in Table~\ref{tab:all-sky}, and detailed in Section \ref{sec:survstrat} below. VLASS uses the VLA to capture the radio spectrum from 2~GHz to 4~GHz in 2~MHz channels, with calibrated polarimetry in Stokes I, Q and U, providing
wideband spectral and polarimetric data for a myriad of targets and source types, thereby addressing a broad range of scientific questions.  VLASS is to be carried out in three passes, each separated by approximately 32 months, resulting in  
a synoptic view of the dynamic radio sky similar to those now available at other wavelengths. VLASS provides
measurements of the radio sky at epochs approximately 20 years after FIRST and NVSS, at sensitivity levels between those of FIRST/NVSS and the upcoming radio surveys from the SKA and its precursors. This is critical to enable early identification and filtering for the most interesting transient events. VLASS will be the first radio survey to image the sky at an angular resolution comparable to those of optical and near-infrared surveys. This resolution does, however, come at a cost. The rms surface brightness sensitivity of the standard image products will be limited to a brightness temperature of $\approx 1.5$\,K, and sources with diffuse emission on angular scales $\stackrel{>}{_{\sim}}$\,30~arcseconds will be poorly imaged and will have their total flux densities underestimated due to lack of short baseline coverage.

The approach adopted provides both astronomers and citizen scientists \citep[e.g.,][]{2015MNRAS.453.2326B} with information for every accessible point of the radio sky. In conjunction with a well-designed archive database and data access tools, VLASS users will be able to implement advanced machine learning techniques to study and classify millions of radio sources \citep[e.g.,][]{2017ApJS..230...20A,2018MNRAS.478.5547A,2018MNRAS.480.2085A,2018MNRAS.476..246L}. 

\subsection{This paper}
In this paper, we describe the science motivation for \hbox{VLASS}, summarize the technical implementation, and show some early results as a resource for potential users of the survey data. Further 
details of the survey may be found in the VLASS memo series, located at \url{https://go.nrao.edu/vlass-wiki}. 
Section \ref{sec:science} describes the science goals of the survey. 
Section \ref{sec:survstrat} describes the survey strategy picked to address those
science goals. Section \ref{sec:impl} outlines the survey implementation plan, with Section \ref{sec:observ} describing the observing plan, and Section \ref{sec:datatech} the data products. Section \ref{sec:commensaldata} describes the commensal surveys that will be carried out at the VLA in conjunction with VLASS. In Section \ref{sec:earlycal}, we describe the calibration of the survey data. (We defer the description of the imaging to memos and future publications, as some imaging algorithms remain under development.) In 
Section \ref{sec:EPO} we describe the Education and Public Outreach efforts aligned with the Survey.  Section \ref{sec:summary} provides a short summary.

\section{VLASS Science Themes} \label{sec:science}

There are four themes that run throughout the VLASS design and science goals. 
These exploit the unique capabilities of the Karl G. Jansky VLA:
\begin{enumerate}
    \item {Hidden Explosions and Transient Events}
    \item {Faraday Tomography of The Magnetic Sky}
    \item { Imaging Galaxies through Time and Space}
    \item { The New Milky Way Galaxy}
\end{enumerate}

\subsection{Key Science Theme 1: Hidden Explosions and Transient Events} \label{sec:transsci}
Radio transients arise from diverse phenomena, including astrophysical blast waves, catastrophic collapses, compact object mergers accompanied by gravitational waves, magnetic acceleration of relativistic charged particles, shocks in high energy particle jets and in magnetized diffuse interstellar plasma, flaring reconnection events in the atmospheres of low mass stars, and the cosmic beacons of rotating neutron stars, white dwarfs, and black hole accretion disks. While the detection of such events has previously relied on the synoptic survey capability of modern wide-field optical, X-ray and $\gamma$-ray observatories, VLASS now also offers the potential to systematically characterize the dynamic sky at radio frequencies, targeting both Galactic and extragalactic transient populations. As described below, the most numerous radio transients, with the greatest potential impact, belong to populations that can be hidden from view in other wavebands, and in some cases are detectable only at radio wavelengths.

VLASS observations will enable searches for transients in multiple ways. Slow transients, with durations from weeks to years, can be found using the VLASS basic data products (Section \ref{sec:basicdata}, in particular the Quick Look images). Fast transients (milliseconds to seconds duration), such as Fast Radio Bursts, will be found using two commensal projects described in Section \ref{sec:commensaldata}. {\em realfast} will search the raw VLASS data stream for transient events, and the VLA Low-band Ionosphere and Transient Experiment (VLITE) will pass its low frequency data through a special pipeline designed to find fast transient events. In the remainder of this Section, we focus on the slow transient population.

\subsubsection{What is the true rate of explosions in the local Universe? }
\begin{figure*}
\begin{center}
\includegraphics[width=15cm]{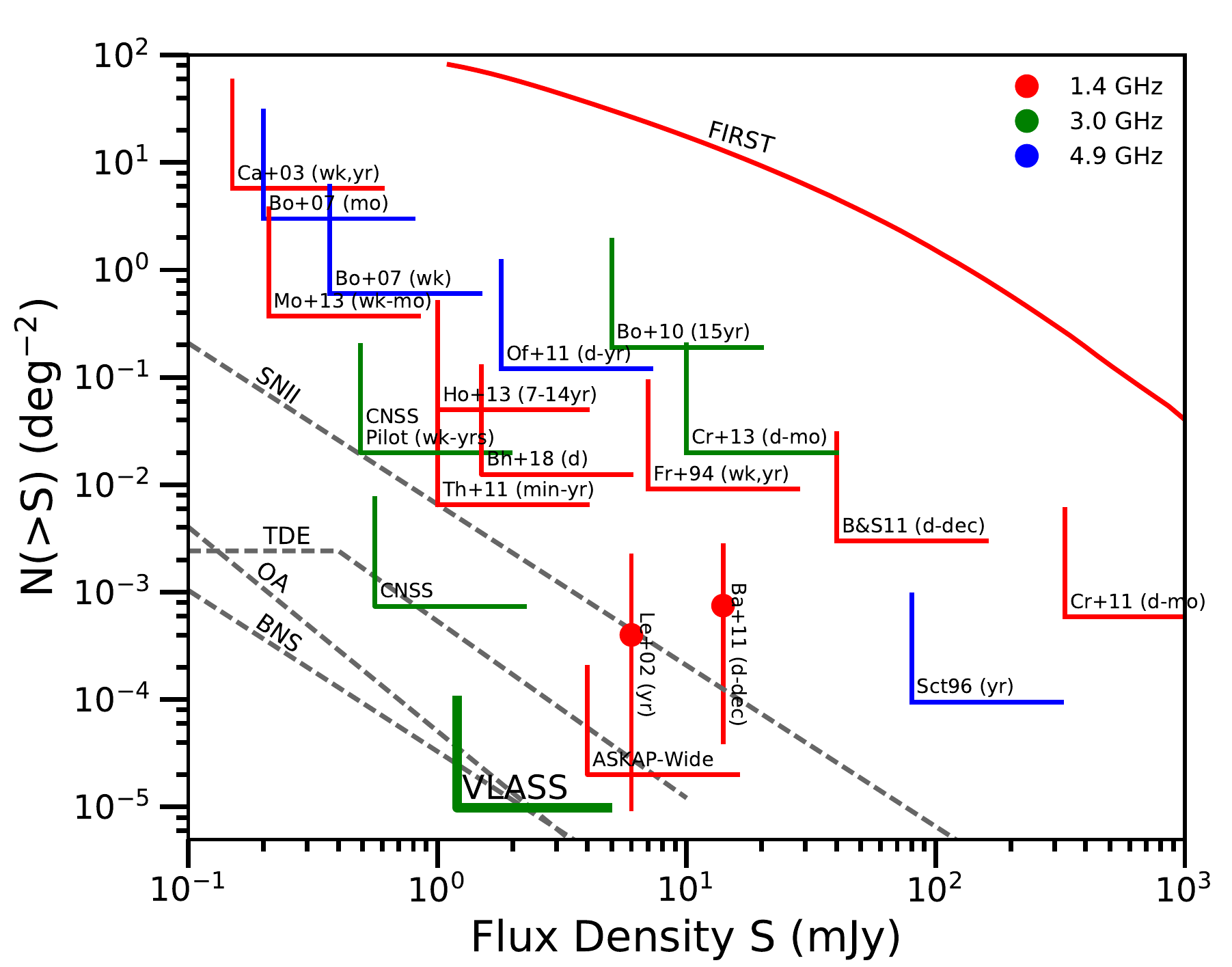}
\caption{\footnotesize The areal densities of extragalactic radio transients. The dashed gray lines give the instantaneous source counts for different classes of transients at 3\,GHz, assuming luminosities and densities from \cite{2016ApJ...818..105M}. The solid red line represents the total 1.4~GHz source counts from FIRST \citep{first95}.  
Wedges indicate upper limits to the transient rates from previous surveys and errorbars (2$\sigma$) are transient rates for past detections (see compilation at \texttt{http://www.tauceti.caltech.edu/kunal/radio-transient-surveys/index.html} for updates). The markers are color-coded according to observing frequency, and indicate the time range to which each survey was sensitive (days (d), weeks (wk), months (mo) years (yr) and decades (dec)). Adapted from \cite{2016ApJ...818..105M}. References: Ba+11, \cite{2011MNRAS.412..634B};Bo+07, \cite{2007ApJ...666..346B}; Bo+10, \cite{2010ApJ...725.1792B}; B\&S 11 \cite{2011ApJ...728L..14B}; Ca+03, \cite{2003ApJ...590..192C}; Cr+11, \cite{2011ApJ...731...34C}; Cr+13, \cite{2013ApJ...762...93C}; Fr+94, \cite{1994ApJ...437..781F}; Ho+13, \cite{2013ApJ...769..125H}; Le+02, \cite{2002ApJ...576..923L}; Mo+13 \cite{2013ApJ...768..165M}; CNSS Pilot, CNSS, \cite{2016ApJ...818..105M};  Of+11, \cite{2011ApJ...740...65O}; Sct96, W.K. Scott, 1996 PhD Thesis, University of British Columbia; Th+11, \cite{2011ApJ...742...49T} 
\label{fig:trans_exgal}
}
\end{center}
\end{figure*}

Phenomena associated with the explosive death throes of massive stars have been studied for decades at all wavelengths. 
These phenomena include the various classes of supernovae, the highly relativistic outflows of $\gamma$-ray bursts (GRBs) and the mergers of compact objects. Also related to these are the tidal disruption events (TDEs) that occur when a star falls into a supermassive black hole, an example of which has recently been identified in a radio survey \citep{2019arXiv191011912A}.
In Figure \ref{fig:trans_exgal} we show a plot of areal density of transients versus their flux density at 3~GHz. The expected rates of different transient classes, shown as dashed gray lines in the plot, are compilations from literature \citep[see references given by][]{2016ApJ...818..105M} (see also \citet{2015ApJ...806..224M} for an independent study that yields similar estimates, of order a few tens each in VLASS of binary neutron star mergers (BNS), GRBs and Orphan GRB Afterglows (OAs), and TDEs).
The limits shown for VLASS represent the expected cumulative number of detected transients, with the requirement that a transient event must be detected at $> 10\sigma$ in a single epoch, to overcome noise fluctuations in the data when comparing two epochs, each with $5\times 10^{10}$ synthesized beams, as well as image artifacts that masquerade as transients. (We note that this is conservative -- one can use much lower detection threshold (e.g., $6\sigma$) combined with the requirement that the detected transient must be located in a nearby galaxy ($z < 0.1$) to overcome high-$\sigma$ thermal noise fluctuations.)

As well as providing a window into the life cycles of massive stars, these explosive events also probe the formation of compact objects and can provide a standard candle for precision cosmology. However, the true rate of such events is poorly constrained; specifically: 

\begin{itemize}

\item A comparison of the star formation rate and supernova (SN) discovery implies that as many as 40\% of SNs remain undetected in the traditional optical searches, largely due to extinction via dust obscuration, with far reaching consequences for models of stellar and galaxy evolution \citep{2012ApJ...756..111M}. 

\item The relativistic outflows associated with GRBs and compact object mergers are highly collimated. Thus, only those bursts that are collimated in the direction of Earth are detected at $\gamma$-ray or X-ray wavelengths. Best estimates suggest this corresponds to a small fraction of the true event rate, dependent on the typical opening angle of the collimated jet. GRBs whose outflows are not pointed towards Earth, but whose unbeamed emission can be detected at cm-radio radio wavelengths are OAs, and can be detected in surveys for radio transients \citep{1997ApJ...487L...1R, 2014PASA...31...22G}.

\end{itemize}

At its reference frequency of 3~GHz, the light curves from extragalactic explosive events can have a characteristic rise time as long as one year \citep[e.g., SNs,][]{2013MNRAS.428.1207S}, and thus the three epochs of VLASS will essentially be three {\em independent} epochs, thereby maximally leveraging the observing time of the VLASS towards the identification of independent transient events. The sensitivity of VLASS to long-duration transients does not come at the expense of sensitivity to transients on shorter timescales, including, but not limited to, Galactic transients and relativistic transients viewed on-axis or close to on-axis. The rapid release of VLASS Quick Look images (Section \ref{sec:basicdata}) will enable early identification of candidate transients whether they are short or long duration events.

Confirmation and classification of VLASS transients will rely on follow-up observations with the VLA, as well as triggered follow-up from the community at optical and NIR wavelengths. The former will sample the light curve of the event and, more importantly, will allow confirmation of the broad Spectral Energy Distribution (SED) of the radio transient (1--50~GHz), a capability unique to the VLA among radio instruments capable of synoptic surveys \citep[at least until the construction of the ngVLA; ][]{2018ASPC..517...15S}. The triggering of optical/NIR follow-up allows detailed characterization of the host, which is the primary tool in classification of a radio transient. Confirmation of distance, and thus energetics, together with the radio transient SED and location within a galaxy, will provide a means to distinguish between the various classes of explosive events \citep[cf.\ Figure 4 of][]{2018ApJ...857..143M}. The shallow depth of individual epochs of the VLASS is well suited to this task. At these depths, a transient event should be identified with a known optical/infrared host galaxy and there should be no ``hostless'' events for the known classes discussed below. Furthermore, the high angular resolution of VLASS allows identification of the location \textit{within} the host galaxy.
This discriminator allows us to assess the likelihood that the event is due to AGN variability or a TDE (if located at host galaxy center), or a SN or BNS merger associated with a catastrophic stellar demise. 

The promise of VLASS for discovering transient radio sources was illustrated by the discovery of a decades long transient using multi-epoch radio survey data, including Epoch 1 of VLASS \citep{2018ApJ...866L..22L}. The transient is located about 0\farcs5 from the center of its host galaxy, and may correspond to the first OA to be discovered \citep{2019arXiv190206731M}.

\subsubsection{The Gravitational Wave Era}

Advanced LIGO (aLIGO) and Advanced Virgo (AdV) recently detected the first confirmed binary neutron star coalescence, GW170817, considerably improving our understanding of these events and confirming their role as the forge within which the bulk of the heavy elements are synthesized \citep{2017ApJ...848L..13A}. Radio observations have been key in establishing the mass and energy of the material dynamically ejected during the merger, investigating the presence of a relativistic jet and constraining the environment of a neutron star merger for the first time 
\citep{2017Sci...358.1579H,2017ApJ...848L..21A,2018Natur.554..207M}. The results to date have been surprising, with the bulk of the radio emission consistent with a wide-angle outflow of mildly relativistic material \citep{2017ApJ...843L..34K,2018Natur.554..207M,2018ApJ...856L..18M,2018ApJ...867...95H,2018MNRAS.478L..18T}, and late-time emission dominated by a narrow energetic jet \citep{2018Natur.561..355M,2018ApJ...868L..11M,2018arXiv180800469G}. 
This event is just the first of an impending torrent of candidate neutron star-mergers. Advanced LIGO and Advanced Virgo resumed operations in April 
2019 (during VLASS observations of the second half of the sky for the first epoch) with a greatly enhanced sensitivity. The sensitivity will increase in subsequent science runs as detectors improve and with the eventual operation of a LIGO detector in India and the KAGRA interferometer in Japan. VLASS will deliver reference image data over most of the sky to help constrain follow-up observations of potential gravitational wave events.

\subsection{Key Science Theme 2: Faraday Tomography of The Magnetic Sky} \label{sec:polsci}

The VLA's WIde-band Digital ARchitecture (WIDAR) correlator has opened a major new window for wideband polarization work, enabling us to characterize properties of the magneto-ionic medium in AGNs and in galaxies across a wide range of redshifts both directly, from their broadband polarized emission, and also when seen against polarized background sources \citep{2014arXiv1401.1875M}. The increase in the sensitivity and the wideband capability of the VLASS will provide a six-fold increase in the density of polarized sources with reliable Faraday rotation compared to what is available today \citep{2009ApJ...702.1230T}.

\subsubsection{Polarization in VLASS}

\begin{figure*}[!t]
\begin{center}
\includegraphics[width=18cm]{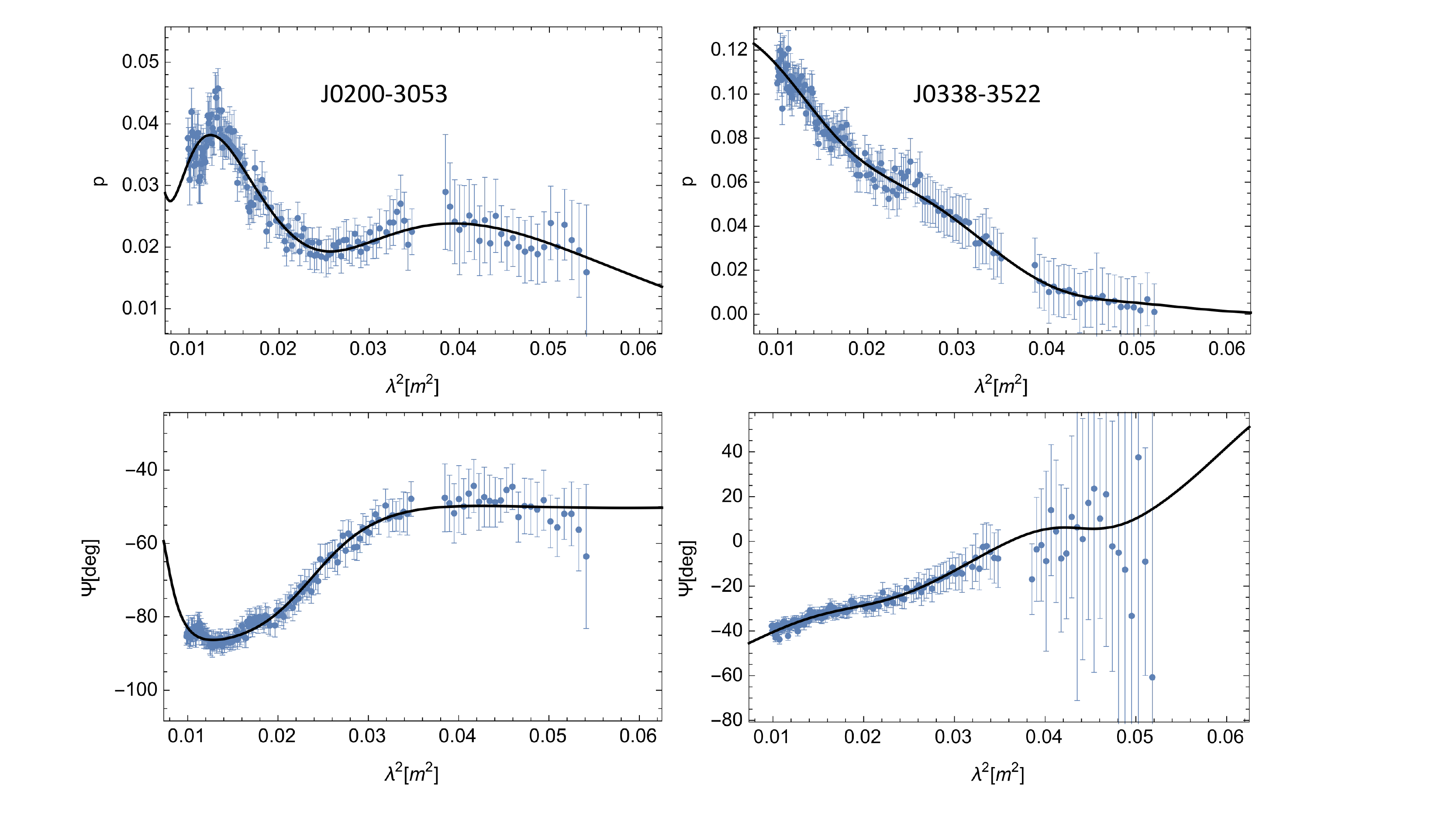}
\caption{Two radio sources with complicated Faraday structures from the Australia Telescope Compact Array observations of \citet{2017MNRAS.469.4034O}. Both the degree of polarization, $p$, and the position angle, $\psi$,
vary rapidly with frequency. Such sources would be possible to identify from VLASS data 
alone (which spans the range $0.006<\lambda^{2}<0.023\, {\rm m^2}$).
}
\label{fig:faradaycomplex}
\end{center}
\end{figure*}

Faraday rotation in a magneto-ionic medium produces various external
or internal depolarization processes
\cite[e.g.,][]{1966MNRAS.133...67B, 1991MNRAS.250..726T,
1998MNRAS.299..189S}.  These provide a unique and critical
diagnostic of the magneto-ionic medium, but only when observed over a
wide frequency range with near-continuous frequency coverage
\citep[e.g.,][Figure \ref{fig:faradaycomplex}]{2012MNRAS.421.3300O,2016ApJ...825...59A,2017Galax...5...66M,
2018arXiv180109731P}. Until a few years ago, polarimetric studies have
either relied on a small number of widely-spaced narrow bands, or
have observed over a continuous but relatively narrow ($\sim$10\%) fractional 
bandwidth. Both these approaches have severe shortcomings. Degeneracies between 
different types of depolarization behavior, and hence the underlying physical
properties of polarized sources and foreground gas, can only be
broken by wideband spectro-polarimetry \citep{2011AJ....141..191F}.

The 2--4\,GHz band of VLASS presents a unique opportunity to examine sources that would have been severely depolarized at lower frequencies. A
useful comparison can be made with the bandwidth depolarization
present in the NVSS survey. The original catalog \citep{nvss} reported band-averaged polarization in two 50~MHz channels centered around 1.4~GHz. 
Therefore, sources with rotation measure (RM) of $|$RM$|>$~100 rad\,m$^{-2}$ were significantly depolarized. 
The \citet{2009ApJ...702.1230T} rotation measure catalog performed a split-band analysis to derive RMs, and thus it is sensitive to a maximum $|$RM$|$ of approximately 500~rad\,m$^{-2}$.  By contrast, the 
higher frequency and smaller channel widths of VLASS mean that the sensitivity of the VLASS will extend to $|$max(RM)$|\approx$16,000~rad\,m$^{-2}$ using 16\,MHz channels.

Even higher maximum rotation measures could be probed by going to the full 2\,MHz native resolution of the VLASS data.  
The frequency coverage sets the resolution in Faraday depth
space, as shown in Figure~\ref{fig:faraday} of Section~\ref{sec:basicdata}, where the width of
the Faraday depth response function is $\approx$200~rad\,m$^{-2}$. For example, 
Faraday components with separations of
$>$100~rad\,m$^{-2}$ would depolarize the NVSS, but with the
VLASS transfer function of 200~rad\,m$^{-2}$ they can be separated and measured 
\citep[e.g.,][]{2015AJ....149...60S}. This provides a resolution of 10~rad\,m$^{-2}$ 
at a signal-to-noise ratio (SNR) of 10, sufficiently small to enable foreground screen experiments, such as the one described in Section \ref{sec:mgii} below.

The 70\,$\muup$Jy rms for the full, combined, three epoch survey will result in $\approx$six polarized sources
per square degree at SNR$=$10, using the 20~cm results of
\citet{2014ApJ...785...45R} and \citet{2014ApJ...787...99S}.  The loss
in polarized intensity at the 2--4~GHz band due to a typical spectral index of $-$0.7 for
the total intensity spectrum
is compensated by the reduced typical wavelength-dependent 
depolarization (by a factor of $\sim$1.3) in the 2--4~GHz band compared with the lower frequency, 1--2~GHz (20~cm) band
\citep{2016arXiv160704914L}.  Note that this result is based on the
9~arcminute resolution SPASS survey \citep{2013Natur.493...66C},
so the reduction in beam depolarization may result in even larger fractional polarizations at 2--4~GHz, $\sim 10$\% (see Section \ref{sec:obspol}).
Thus, the VLASS sensitivity gives at least a factor of six times
the polarized source density of the \hbox{NVSS}, a major improvement for
targeted foreground experiments,  and will, in addition, allow
much greater access to the rare population of heavily depolarized sources.

Foreground experiments are limited by the scatter in RM in the background sources.
A fundamental limit
to the RM scatter comes from the intrinsic variation in the
extragalactic sources themselves, currently estimated to be $\sim$5--7
rad\,m$^{-2}$ \citep{2010MNRAS.409L..99S}. The VLASS can approach
this limit for polarized flux densities $\gtrsim$1.5~mJy, providing a
source density of $\sim$3 per square degree, an order of magnitude
higher than the NVSS. 

Based on the counts from NVSS, over $10^5$ sources with polarized fluxes
$>$0.75~mJy (SNR$>$10; 5\% [0.5\%] for I~=~15~[150]~mJy) will be detected in VLASS. VLASS will thus enable the first large depolarization catalog to be produced, from which the origins of the depolarization can be statistically evaluated.

The relatively
high frequencies of the VLASS 2--4~GHz polarization survey will also
uncover previously unknown populations of sources with extremely large
Faraday depths and those that are heavily depolarized due to Faraday
effects, and hence missed in the lower frequency catalogs. 

\subsubsection{The magneto-ionic medium in AGN, galaxies and
their immediate environments} 

As described in Section \ref{sec:feedback},
feedback from AGN is important in galaxy formation. Thus, investigating how radio galaxies impart energy into the ISM/IGM is crucial.
While minimal interaction between radio lobes and the environment
would lead to a thin ``skin'' of thermal material around the lobes
\citep[e.g.,][]{1990ApJ...357..373B,2018MNRAS.476.1596K}, significant interaction should
lead to large-scale mixing of thermal gas with the synchrotron
emitting material throughout the lobe, causing internal Faraday
dispersion \citep[e.g.,][]{2016ApJ...825...59A}. \citet{2013ApJ...764..162O} fitted the
depolarization trend of the lobes in one such radio galaxy,
Centaurus~A, and found a thermal gas of density 10$^{-4}$ cm$^{-3}$
well mixed with synchrotron-emitting gas in the lobes. A sensitive
wide-band polarization survey allows statistical studies of this
phenomenon through estimation of the thermal gas content in a large
number of radio galaxies, covering a range of luminosities, redshifts,
and environments. 
Most of these polarized sources will be spatially resolved by the VLASS \citep{2014ApJ...785...45R}, allowing, to first
order, the separation of depolarization along the line of sight from beam depolarization (across the line of sight).

\subsubsection{The emergence and growth of large-scale magnetic fields in galaxies} 

Spatially resolved images of the polarized synchrotron
emission from nearby galaxies demonstrate the existence of ordered magnetic fields 
in the ISM \citep[e.g.,][]{1996ARA&A..34..155B,2015A&ARv..24....4B}.  However, the
evolution of galactic-scale magnetic fields over cosmic time is poorly
constrained, because this traditional approach becomes increasingly
challenging for distant galaxies. An alternative approach is to
utilize the statistics of integrated synchrotron polarization of
unresolved galaxies to infer their overall magnetic field properties
\citep[e.g.,][]{2009ApJ...693.1392S}.  In the presence of a
large-scale galactic field, the position angle of the integrated
polarized radiation is aligned with the minor axis of the galaxy for
rest frame frequencies above a few GHz.

In star-forming galaxies, which are mostly spirals, polarization reflects the degree of ordering in the
intrinsic disk field. The fractional polarization distribution of nearby disk galaxies at 4.8 GHz was measured by \cite{2009ApJ...693.1392S} and \cite{Mitchell2009}, who carried out a survey of 47 nearby (within 100 Mpc) galaxies using the Effelsberg Telescope.  These data show that at least 60\% of normal spiral galaxies show a degree of polarization higher than 1\%, and in some cases higher than 10\%. Moreover, there is a strong correlation between polarization position angle and the optical minor axis of the galaxy disk. Unlike the Effelsberg observations, VLASS resolves such galaxies, reducing beam depolarization, and thus we expect to detect even higher fractional polarizations.

\subsubsection{Quasar absorption line systems}\label{sec:mgii} 
Mg II absorbers are
associated with a $\sim$10$^4$~K photoionized circumgalactic medium in
a wide range of host galaxy types and redshifts \citep[see][for a
  review]{2005ASPC..331..387C}. These systems potentially trace
outflows from star formation \citep[e.g.,][]{1996ApJ...472...73N} and
cold-mode accretion \citep[e.g.,][]{2010ApJ...711..533K}.  When seen
against polarized background sources, the Faraday depth provides a
direct measure of the electron density and the magnetic field strength
in Mg II absorbing systems, parameters that are both currently poorly
constrained. \citet{2008Natur.454..302B, 2013ApJ...772L..28B}, \citet{2013MNRAS.434.3566J}, \citet{2016ApJ...829..133K} and \citet{2014arXiv1406.2526F,2017ApJ...841...67F} 
have demonstrated the presence of larger
RM scatter in systems associated with Mg II absorbers, and have
interpreted this as evidence for magnetic fields with strengths $\sim 1 \mu$G out to $z\sim2$, possibly
associated with outflows. Polarization data from VLASS will be able to significantly
increase the sample size of such studies from the $<100$ currently available, and rigorously
test models for circumgalactic gas and
photoionization, and the evolution of large-scale
magnetic fields over cosmic time \citep{2018MNRAS.477.2528B}.

\subsubsection{Spatially resolving magnetic field structures in distant galaxies}
Strong gravitational lensing of polarized background quasars by foreground galaxies has proven to be a clean and effective way of measuring the in situ magnetic fields in individual cosmologically distant galaxies. \citet{2017NatAs...1..621M} used the broadband polarization properties of the lensing system CLASS B1152+199 observed with the VLA and derived coherent axisymmetric magnetic fields at the $\mu$G level in the star forming lensing galaxy at $z=0.44$, currently the most distant galaxy for which we have both a magnetic field strength and a geometry measurement. 
Wavelength-dependent depolarization towards the lensing system can also provide information on turbulence in the magnetized medium in distant galaxies. A crude estimation based on the the statistics of the Cosmic Lens All-Sky survey \citep[CLASS;][]{2003MNRAS.341....1M} suggests that the VLASS has the potential to identify $\sim 10$ new wide separation ($>2\farcs 5$) gravitational lensing systems.

\subsection{Key Science Theme 3: Imaging Galaxies through Time and Space}

\subsubsection{The Evolution of Accretion Activity in Active Galactic Nuclei}\label{sec:agn}

There is now strong evidence that the standard Active Galactic Nucleus (AGN) unification paradigm \citep[e.g.,][]{1993ARA&A..31..473A, 1995PASP..107..803U} is incomplete.
For example, observational evidence \citep[e.g.,][]{2007MNRAS.376.1849H, 2010MNRAS.406.1841H, 2012MNRAS.421.1569B} suggests that many or most low-luminosity ($L_{\rm 1.4~GHz} < 10^{25}$\,W\,Hz$^{-1}$) radio galaxies in the local universe correspond to a distinct type of AGN. These sources accrete through a radiatively inefficient mode (the so-called ``radio mode"), rather than the radiatively efficient accretion mode typical of radio-quiet optically or X-ray selected AGN [sometimes called ``quasar mode"; see \citet{2014ARA&A..52..589H} for a recent review covering these feedback processes]. The role of these two accretion modes appears to be strongly influenced by the environment \citep[e.g.,][]{2008A&A...490..893T} while the presence or absence of a radio-loud AGN appears to be a strong function of the stellar mass of the host galaxy \citep[e.g.,][]{2005MNRAS.362...25B, 2012A&A...541A..62J, 2015MNRAS.450.1538W}.

The combination of sensitivity and angular resolution of VLASS allows the study of the entire AGN population from classical radio-loud sources down to the realm of radio-quiet AGN at low redshifts \citep[$L_{\rm 1.4~GHz}\sim$10$^{22-23}$\,W\,Hz$^{-1}$;][]{2004NewAR..48.1173J, 2008MNRAS.388.1335W, Kimball11, 2013ApJ...768...37C}, and, with stacking, out to significantly higher redshifts \citep[e.g.][]{2007ApJ...654...99W,2017MNRAS.466..921C}. VLASS is sensitive enough to detect luminous radio galaxies and radio-loud quasars out to $z\sim 5$ \citep[typical 3~GHz flux densities of a few mJy, e.g.,][]{1999ApJ...518L..61V, 2018arXiv180601191S, 2018ApJ...861L..14B}, allowing the study of their evolution from the first billion years of the Universe to the present day.  
Morphology and spectral index information from VLASS will allow the evolution of different morphological (e.g.\ FRI, FRII) and physical (e.g.\ compact steep spectrum, GHz-peaked spectrum, blazar and BLLac) classes of radio-loud AGN to be quantified.

\begin{figure}[!t]
\centering
\includegraphics[width=0.5\textwidth]{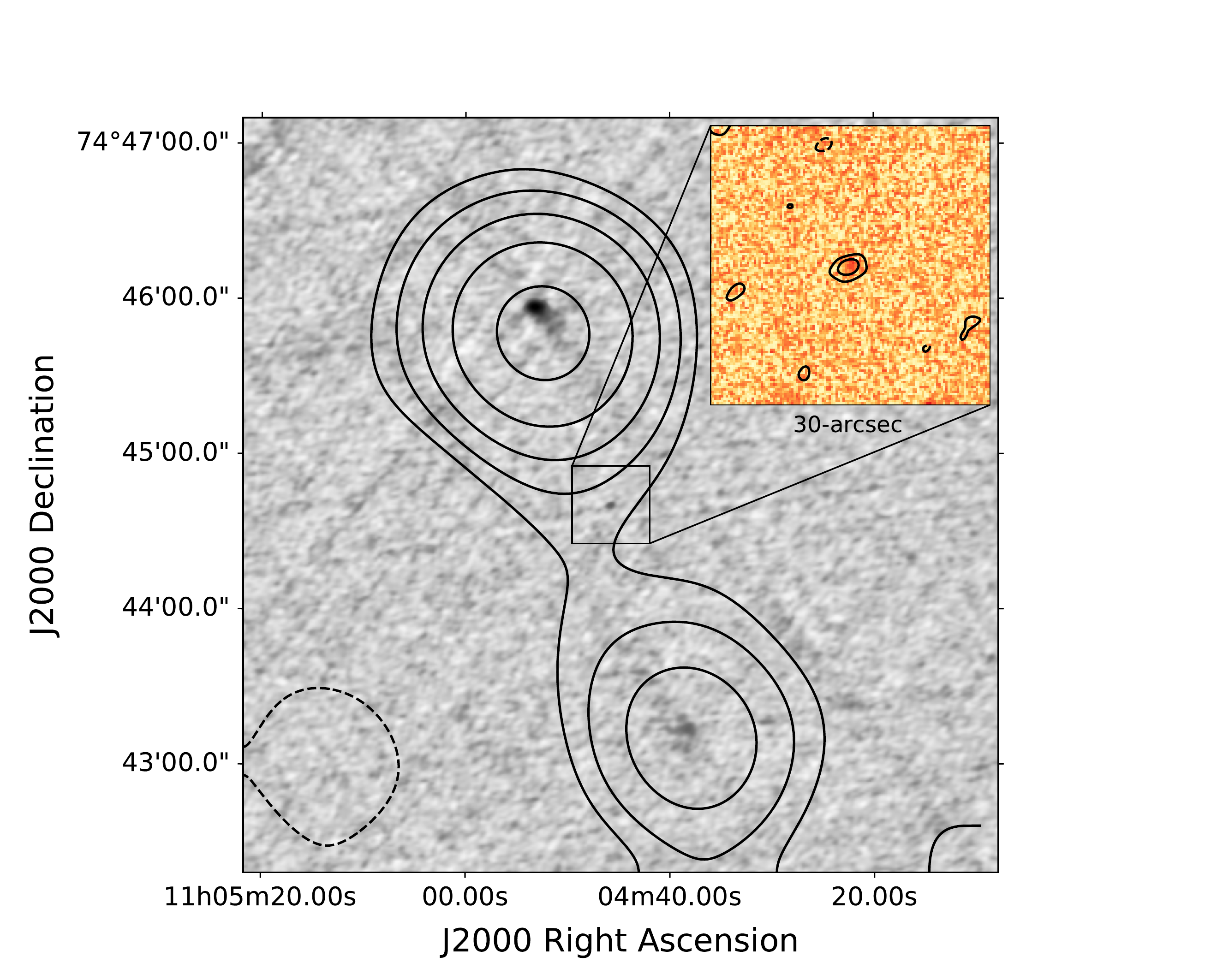} 
\caption{An example of a giant radio galaxy whose host galaxy has been identified using VLASS data. The greyscale shows the VLASS image, the contours are from NVSS (at -0.75, 0.75, 1.5, 3.0, 6.0 and 12 mJy/beam). The inset shows the PanSTARRS $i$-band image with VLASS contours at -0.26, 0.26 and 0.52 mJy/beam.} 
\label{fig:giant}
\end{figure}

\subsubsection{AGN Feedback}\label{sec:feedback}

Feedback from AGN on the ISM of their host galaxies is important in
galaxy formation: it is intimately linked to the star formation
history \citep[e.g.,][]{2006ApJ...651..142H}, and could suppress
cooling in massive galaxies, producing the bright-end cut-off of the
luminosity function \citep[e.g.,][]{2006MNRAS.368L..67B,
  2006MNRAS.365...11C}. The nature of this AGN feedback is very much
under debate: it has been shown that energy deposited by radio jets
can either trigger or quench star formation \citep[e.g.,][]{2012ApJ...757..136W}. Recent studies from both a theoretical \citep{2013ApJ...772..112S} and observational \citep{2012MNRAS.427.2401K, 2014MNRAS.442.1181K} perspective have shown that powerful radio-loud AGN may actually provide a positive form of feedback. On the other hand, there is little evidence for any type of feedback from radio-quiet objects based on the latest studies using \textit{Herschel} \citep[e.g.,][]{2011MNRAS.416...13B, 2013A&A...560A..72R,2016MNRAS.462.4067P}. Moreover, the interplay between jets and feedback in nearby satellite galaxies is even more poorly understood \citep[e.g.,][]{2006ApJ...647.1040C,2017ApJ...838..146L}. 

The morphological information from VLASS allows the selection of large samples of specific types of radio source to 
aid the understanding of feedback. For example, extended FRI sources for ``radio mode'' feedback in groups and clusters
\citep[e.g.,][]{2012A&ARv..20...54F,2014arXiv1401.0329C,2017MNRAS.472.4024R} (visible out to $z\approx 0.5$), and powerful, compact sources for ``quasar mode'' feedback and interactions between jets and ionized gas in compact radio galaxies \citep[e.g.,][]{1996AJ....112..902S, 2012MNRAS.424.1346W, 2015ApJ...813...45L} (visible out to $z>2$) can be identified.

\subsubsection{Radio source environments}

The details of the mechanism(s) of interaction between radio-loud AGN and their environments, on all scales, remain unclear. Basic questions, such as whether the most powerful sources are expanding supersonically throughout their lifetimes \citep[e.g.,][]{1989ApJ...345L..21B, 2000MNRAS.319..562H}, or what provides the pressure supporting the lobes of low-power objects \citep[e.g.,][]{2008ApJ...686..859B, 2008A&A...487..431C} remain unanswered. These questions can only be addressed by the accumulation of large, statistically complete samples of radio sources with imaging capable of resolving them on scales $\sim$10~kpc, combined with excellent multi-wavelength data. Information on both large and small-scale radio structure is required. As an illustration, \citet{2018arXiv180807178V} use VLASS Epoch 1 data to discover new giant ($>1$~Mpc) radio sources. VLASS can detect both the extended lobes and the 
unresolved cores of these objects, which can then 
be identified with the correct host galaxy (e.g.\ Figure \ref{fig:giant}).

\subsubsection{Radio emission from AGN}
There is nothing in the optical/UV properties of quasars
(continuum, emission, absorption) that can be used to predict whether a given 
quasar is radio-loud or not \citep[e.g.,][]{Kratzer14}.  
An important avenue forward is to have
better demographics across a broad range of luminosity and redshift, 
seeking to confirm marginal evidence that objects are more likely to
be strong radio sources at lower redshift and higher luminosity.
Further insight can be gained from radio demographics as a function of quasar emission line ratios and widths
\citep[e.g.,][]{2002ApJ...565...78B}. 

Even radio-quiet quasars are not radio-silent.  It is not clear whether the radio emission is from failed jets, star formation, or shocks \citep[e.g.,][]{Kimball11,Condon13,Zakamska14, 2017MNRAS.468..217W}, indeed  there is evidence that all three may contribute. VLASS will provide better demographics -- both for direct detection (out to $z\approx 0.1$) and stacking analysis (for more distant radio-quiet quasars) -- key to understanding this question.

\subsubsection{Dual AGN}

\begin{figure*}[!t]
\centering
\includegraphics[width=1.0\textwidth]{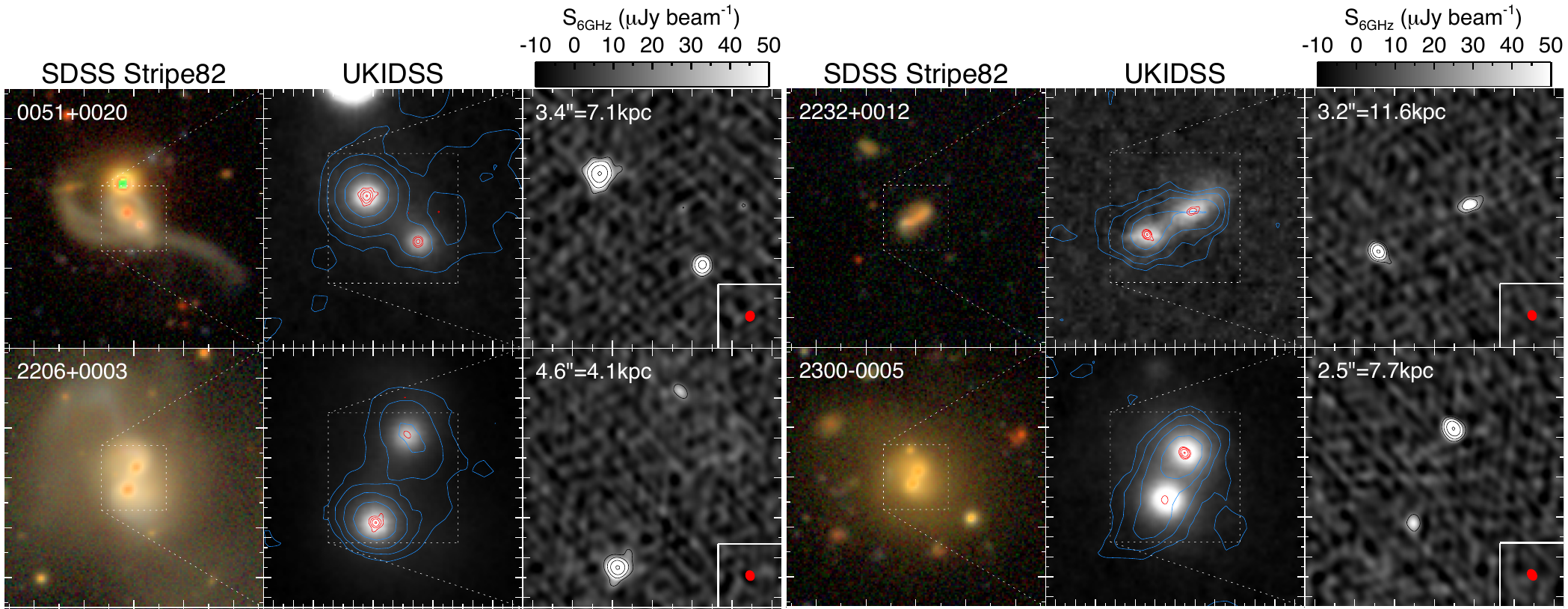} 
\caption{Kpc-scale dual radio AGN mined from the VLA Stripe 82 survey. For each system, we show a wide-field SDSS deep coadded $gri$ color image, a narrow-field UKIDSS $J$-band image overlaid with the 1.4\,GHz contours from VS82 ({\it blue}; 1\farcs8 beam) and the VLA 6\,GHz A-configuration continuum contours ({\it red}; 0\farcs3 beam), and a zoom-in on the VLA 6\,GHz intensity map. The projected separation of each pair is labeled. VLASS will cover more than 300 times more area than VS82, so, despite the lower resolution of VLASS, many examples of dual AGN at $z\sim 0.1$ to $z>1$ with separations $\sim 3$ to $\sim 30\;$kpc will be found.} 
\label{fig:binaries}
\end{figure*}

It has long been suspected that the development of gravitational instabilities in galaxy mergers may drive gas deep to the vicinity of the supermassive black hole and trigger episodes of AGN activity \citep{Shlosman89,Barnes91}. However, the key small-scale physical processes remain poorly constrained by observations and unresolved in simulations. In particular, because of observational challenges, only a few examples are known of the population of late-stage mergers in which the black holes in both host galaxies accrete and power AGN predicted by simulations  \citep[e.g.,][]{Van-Wassenhove12,Blecha13}. At radio wavelengths, an ongoing challenge has been having enough sky coverage covered with sufficient resolution to obtain a substantial number of detections of pairs at separations below a few kiloparsecs.

VLASS is ideally suited to identify a large, uniform sample of sub-galactic-scale dual AGN as it will overcome the three major observational challenges faced in previous surveys by offering: (1) arcsecond-level angular resolution, (2) an enormous survey area combined with sub-mJy sensitivity, and (3) a generic AGN indicator (high surface brightness radio emission) unaffected by dust obscuration \citep[see also][]{2014arXiv1402.0548B}. Previous searches in the 92 square degree VLA Stripe 82 survey \citep[VS82;][]{Hodge11} have shown a high success rate when direct mining of pairs using high-resolution radio data is performed \citep[e.g., Figure~\ref{fig:binaries};][]{Fu15,Fu15b}. 
VLASS has comparable sensitivity and angular resolution to the VS82, making it possible to apply the same technique to a 370$\times$ larger area, dramatically increasing the sample size; the formulation of \citet{2014arXiv1402.0548B} estimates thousands of merging galaxies will be present in the sample. By comparing the large sample of uniformly selected dual AGN with matched control samples, a number of outstanding questions in the merger paradigm of galaxy evolution can be addressed: Do mergers trigger and synchronize black hole accretion? What is the origin of the two accretion/feedback modes (radiative-mode vs.\ jet-mode)? Can AGN feedback significantly affect the star formation activity in Myr timescales? As the immediate precursor of massive black hole binaries, studies of kpc-scale galaxy mergers will also have important implications for low-frequency gravitational wave experiments.

\subsubsection{Star formation in galaxies}

\begin{figure*}[!t]
\centering
\includegraphics[width=0.5\textwidth]{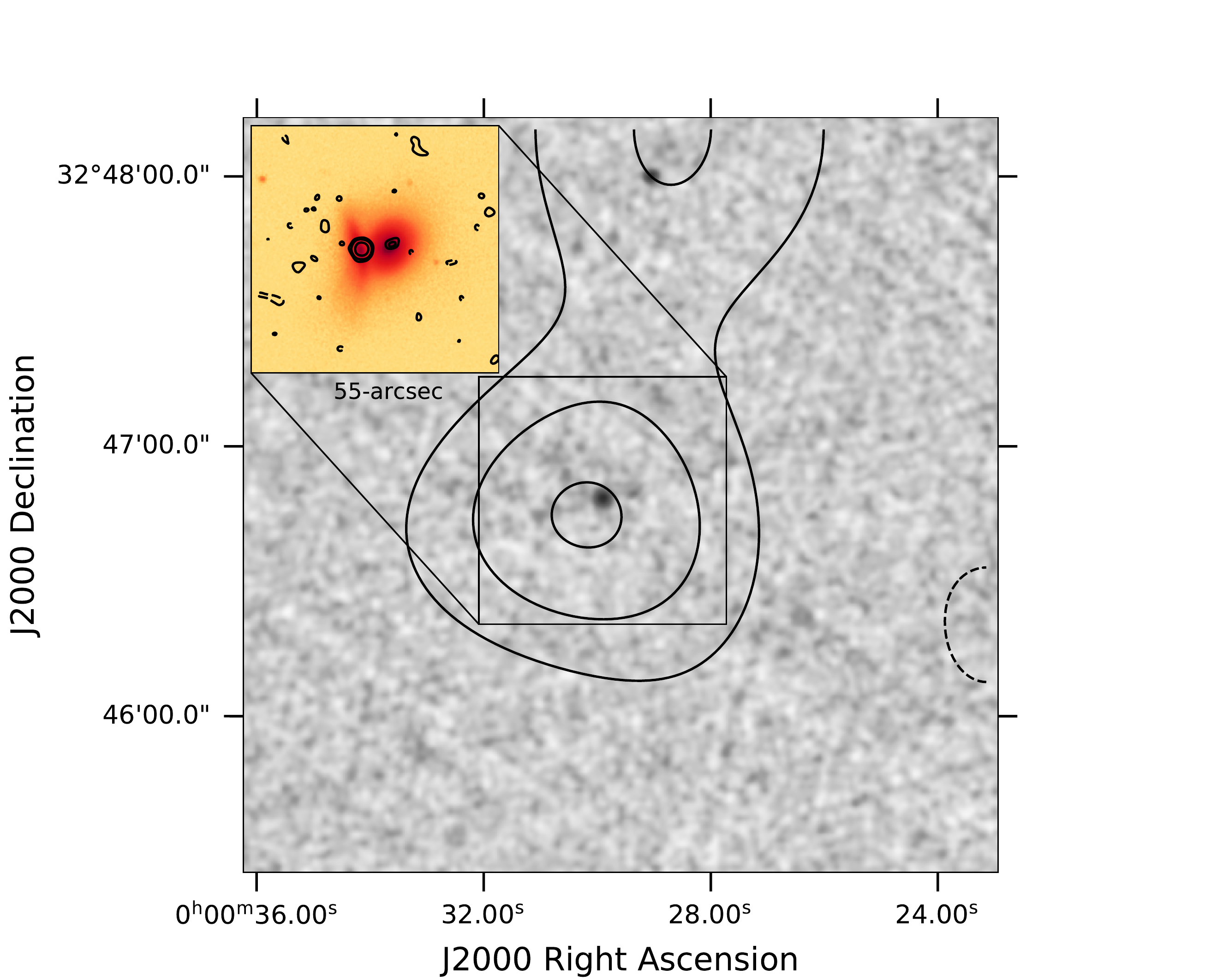}\includegraphics[width=0.5\textwidth]{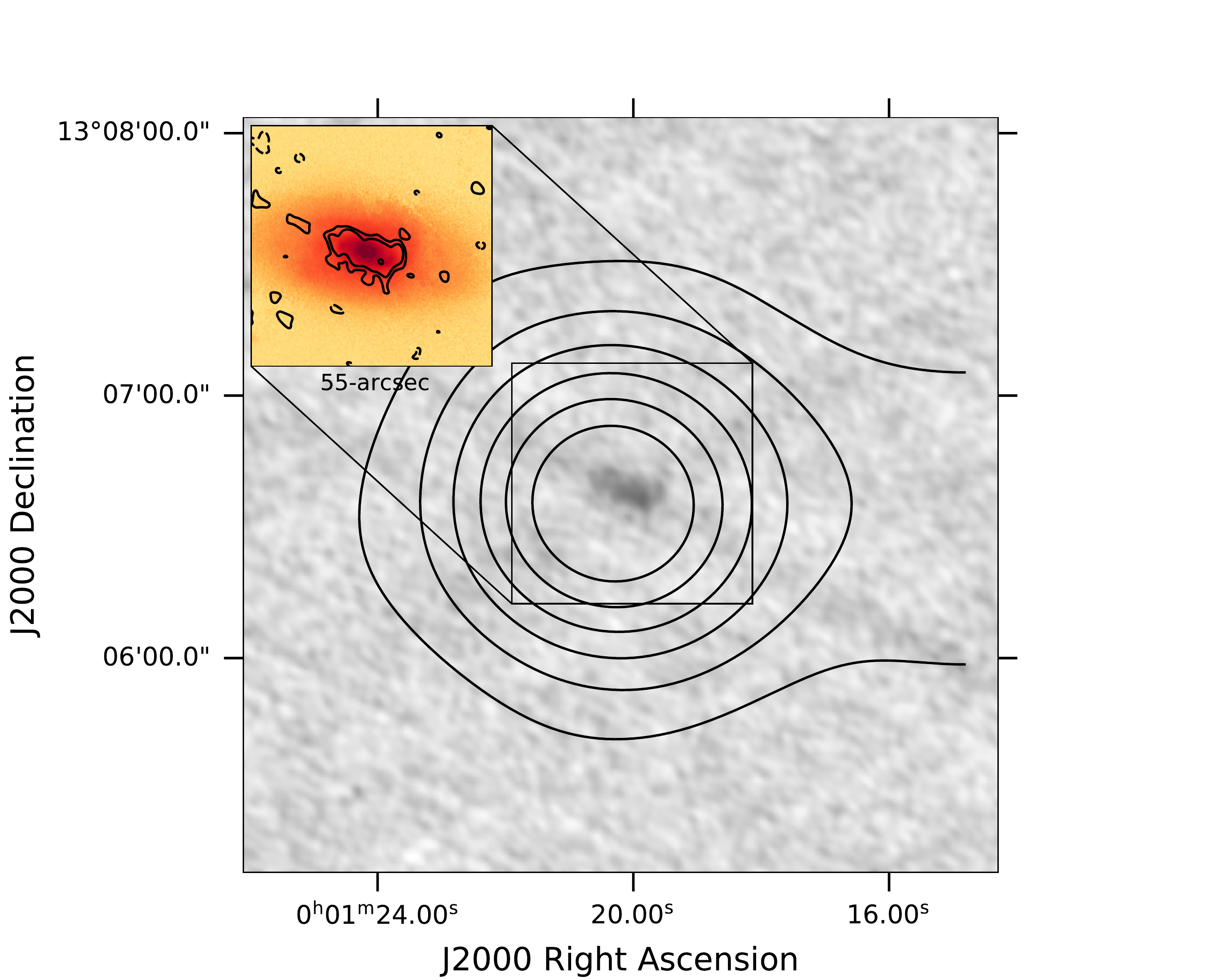} 
\caption{Two nearby starforming galaxies from the sample of \citet{2019ApJ...872..148C} in VLASS. {\em Left}, IC 5373, a merging galaxy pair at $z=0.033$, {\em right} NGC 7803 at $z=0.018$. In both images, the VLASS data are shown in greyscale, and contours are from NVSS (at -1, 1, 2, 3, 4, 5 and 6 mJy\,beam$^{-1}$). The insets show PanSTARRS $r$-band images with VLASS contours superposed (levels -0.26, 0.26, 0.39, 0.78 and 1.56 mJy\,beam$^{-1}$). In both cases, not only are the galaxies detected, but also spatially resolved. In particular, both of the merging pair in IC 5373 are separately detected, with the easternmost galaxy having a compact radio structure that may indicate either a nuclear starburst or a weak AGN.} 
\label{fig:galaxies}
\end{figure*}

The surface brightness sensitivity of VLASS is not high enough to detect diffuse emission far out in galactic disks, even in nearby ($z\lesssim0.1$) systems. However, emission from the higher brightness central regions of normal star forming galaxies is detectable, even in a single epoch of VLASS, and VLASS has the resolution to identify compact star forming regions or weak AGN in the nuclear regions (Figure \ref{fig:galaxies}).

VLASS is deep enough to detect emission from star formation in ultraluminous infrared galaxies (ULIRGS) out to $z\sim 0.5$. This will allow us to increase the number of radio detections by a factor $\approx 5$ compared to studies using FIRST \citep[e.g.,][]{2000ApJS..131..185S}. Furthermore, at VLASS frequencies and lower we expect the radio spectra of the most intense, compact starbursts to flatten or turn over due to thermal absorption by ionized gas \citep[e.g., Arp220;][]{2016A&A...593A..86V}. By measuring the radio spectra of $z<0.5$ ULIRGs using VLASS and other data, we will be able to assess the column density of ionized gas towards the star forming regions, and estimate the free-free emission at high frequencies.

\subsection{Key Science Theme 4: The New Milky Way Galaxy} \label{sec:galsci} 

At visible wavelengths, dust obscuration and absorption effects can be significant toward the Galactic plane, but these effects are not problematic at radio wavelengths, and there is a rich history of using radio observations
to find objects deep within the Galaxy \citep[see][and references therein]{2006AJ....131.2525H}.  Further, there has been a
host of surveys of the Galaxy at wavelengths that do penetrate deep into the Galaxy, from the infrared to $\gamma$-ray (e.g.,
with \textit{Spitzer}, \textit{Herschel}, \textit{Chandra}, \textit{XMM-Newton}, and \textit{Fermi}).
VLASS will cover a wide range of Galactic longitude: $-15^\circ \lesssim \ell \lesssim 260^\circ$ ($\approx 75$\% of the Galactic Plane). VLASS is complementary to an on-going VLA survey, GLOSTAR (A Global View of Star Formation in the Galaxy; \cite{GLOSTAR}) which will map a smaller region of the Galactic Plane at $\mid b \mid \, < 1^\circ$ in 4--8 GHz continuum (to $\approx 40\, \mu$Jy rms) and several molecular and recombination lines. 
VLASS naturally lends itself to finding a variety of Galactic
radio sources. The panoply of Galactic science means that there is a large discovery space for unexpected phenomena. In the remainder of this section, we highlight three topics that will be further illuminated by the Galactic coverage afforded by the VLASS.

\subsubsection{Compact Objects}

The 2--4~GHz frequency of VLASS makes it 
especially sensitive to rare pulsar systems
that are likely to be at low Galactic latitudes close to the Galactic Plane, and therefore  have their emission highly scattered (e.g., PSR-BH binaries). By applying a series of ``filters'' such as spectral index, polarization and compactness, the large number of sources detected in VLASS can be reduced to a feasible number on which to conduct a periodicity search using single dish telescopes \citep[e.g.,][]{2018MNRAS.475..942F}. 
Specific goals for pulsar seaches with VLASS include: the improvement of electron-density and magnetic field models for the Milky Way \citep[e.g.,][]{2002astro.ph..7156C, 2017ApJ...835...29Y,2011ApJ...728...97V, 2012ApJ...761L..11J} by finding more pulsars in the Galactic Plane and Bulge, the discovery of compact binary systems (powerful laboratories for General Relativity), recycled (``millisecond'') pulsars for incorporation into pulsar timing arrays, and exotic systems such as pulsar-black hole systems.

\subsubsection{Coronal Magnetic Activity on Cool Stars}\label{sec:galsci.m}

The radio emission from nearby active stars provides a unique probe of accelerated particles and 
magnetic fields that occur in them, which is useful for a broader understanding of dynamo processes in 
stars, as well as the particle environment around those stars.
The large magnetic field strengths now known to occur around some brown dwarfs were first detected through their effect on cm-wavelength radio emission \citep{2001Natur.410..338B,2006ApJ...653..690H} before the signatures were seen through Zeeman splitting of absorption lines at near infrared wavelengths \citep{2007ApJ...656.1121R}.   
The stellar byproduct of exoplanet transit probes like \textit{Kepler}
and \textit{TESS} will yield information on key stellar parameters like
rotation, white-light flaring, and asteroseismic constraints on
stellar ages.
Unlike other diagnostics of magnetic activity, such as coronal emission, which displays
a maximum value of $L_{X}/L_{\mathrm{ bol}}\approx10^{-3}$,
incoherent stellar radio emission usually shows no saturation effects. Thus,
the level of radio emission can vary by orders of magnitude depending on
the instantaneous particle acceleration events and plasma properties. 
Stellar flares and energetic events in the photospheres, chromospheres and coronae  
of late-type stars are receiving new attention across many wavelength regimes of the electromagnetic spectrum, as the number of known magnetically-active stars with exoplanets increases due to missions like \textit{Kepler} and \textit{TESS}. The impacts of flares, energetic winds, CMEs and other phenomena on the development of exoplanet atmospheres and the habitability of planets may be significant \citep{2007AsBio...7...30T, 2015ApJ...806...41C}.

Scaling from the luminosity distribution of currently known active
stars, \hbox{VLASS} will be able to detect ultracool dwarfs to distances of~10--20~pc,
active dwarf stars to a few tens of parsecs, and active binaries to
slightly less than 2~kpc. 
At the distances to the nearest star forming regions (150--300~pc), \hbox{VLASS}
sensitivity limits will probe stellar radio luminosities
above~1--$4\times 10^{16}$\,erg\,s$^{-1}$\,Hz$^{-1}$, enabling studies of particle acceleration in young stellar objects (YSOs) at a range of evolutionary stage, from
classes I--\hbox{III}.

The combination of VLASS with the LSST and {\em eROSITA} surveys will form a foundation
for the identification and study of nearby active stars, covering the evolution
of magnetic activity from YSOs from the early stages of star formation until they join the main sequence.  
Particularly powerful will be cross-correlating variable or transient sources identified in
LSST surveys with \hbox{VLASS} 
and vice-versa, enabling unbiased constraints on sources producing 
extreme magnetic activity.
The multi-epoch  nature of \hbox{VLASS} will allow for constraints on 
variability of particle acceleration. 
As \cite{2018ApJ...855L...2M} recently demonstrated with their discovery of flares from Proxima Centauri using ALMA, there is still discovery potential for magnetic activity in the radio domain.

\subsubsection{Star Formation and Evolution, Distant Thermal Sources, and Galactic Structure}\label{sec:galsci.pne}
Young, massive stars produce an H\,\textsc{ii} region, while
intermediate and low-mass stars end their lives by expelling their
outer layers into the interstellar medium (ISM) and producing
planetary nebulae (PNe).  In both cases, the hot central star ionizes
the surrounding material, which then emits free-free radio emission.
Radio observations provide a powerful means of identifying these thermal Galactic sources because they are relatively unaffected by dust
obscuration that affects visible and near-infrared wavelengths
\citep[e.g.,][]{2012PASP..124..939H,2014RMxAA..50..203K}.

Typical time scales for an H\,\textsc{ii} region or PNe to evolve are
short.  For example, the expansion time scale of an H\,\textsc{ii}
region may be only $10^5$~yr
\citep{1990ApJ...349..126F,1996ApJ...469..171G}, and it may take only
$10^4$~yr for a star to evolve from the asymptotic giant branch (AGB)
through the PNe phase to become a white dwarf
\citep{2005JKAS...38..271K}. Population synthesis models predict a range in the numbers of Galactic planetary nebulae \citep{2014MNRAS.443.3388S}, but the total number of known nebulae from optical searches is significantly lower than even the most conservative prediction. Consequently, a sensitive, all-sky survey unaffected by dust is required to correctly trace evolutionary sequences.  Further, as
tracers of massive star formation, H\,\textsc{ii} regions are a
natural means to map out the spiral structure of the Galaxy.

Detecting distant thermal sources, such as H\,\textsc{ii} regions and
PNe, requires a combination of both angular resolution and adequate
brightness temperature sensitivity.  An H\,\textsc{ii} region might be
several arcseconds in size (0.5~pc at a distance of 10~kpc), while surveys of nearby
PNe show that they tend to be a few to several arcseconds in size
\citep{1990A&AS...84..229A}.  Given that most H\,\textsc{ii} regions
and PNe are not simple spherical shells, but show more compact
sub-structure, an angular resolution of a few arcseconds is desirable.
Produced by the free-free emission from ionized gas, expected
brightness temperatures might be a few hundreds of degrees to $10^3$~\hbox{K}.  Thus, arcsecond angular resolution and a brightness
temperature sensitivity of order $10$~K or better is sufficient
to detect large numbers of distant thermal sources. VLASS, combined with radio observations at other
frequencies and infrared observations, will provide
an expanded sample of H\,\textsc{ii} regions and PNe 
throughout 75\% of the Galactic disk, allowing a fuller census of these 
phenomena.

\subsection{VLASS in the context of upcoming astronomical facilities}

VLASS stands on the shoulders of the pioneering NVSS and FIRST surveys carried out with the VLA from 1993 to 2011. VLASS additionally sets the stage for future radio surveys and facilities, such as surveys using {\em MeerKAT} (\cite{2016mks..confE...1J} [in particular the MeerKAT International GHz Tiered Extragalactic Exploration Survey (MIGHTEE) and MeerKAT Large Area Sky Survey (MeerKLASS) \citep{2016mks..confE...6J, 2016mks..confE..32S}];
surveys using the {\em Australian Square Kilometer Array Pathfinder} [{\em ASKAP}, in particular the Evolutionary Map of the Universe (EMU) survey \citep{2011PASA...28..215N} and the Polarization Sky Survey of the Universe's Magnetism (POSSUM) \citep{2010AAS...21547013G}]; the {\em Westerbork Synthesis Radio Telescope} [{\em WSRT}/Apertif; in particular the WODAN survey, \cite{2011JApA...32..557R}], and the ongoing {\em LoFAR} Two Metre Sky Survey \citep{2017A&A...598A.104S},
leading to Phase 1 science operations of the {\em Square Kilometre Array (SKA)} that will commence in the 
late 2020s. Each of these facilities includes dedicated surveys at $\approx$1.4\,GHz as a prime component of their science programs \citep{2017NatAs...1..671N}. The observing band and parameters of VLASS, and the fact that VLASS surveys 65\% of the Southern sky, means that it will be complementary to these lower frequency programs. 

The combination of VLASS with SKA precursor surveys at $\approx$1--2~GHz --- MIGHTEE, POSSUM/EMU and WODAN --- will provide 1--4~GHz of frequency coverage to increase the ability to characterize complex spectral and Faraday structures, opening up new and exciting science that combines the strengths of the VLA and SKA precursors.
The high frequency and high angular resolution of the VLA were important drivers in the design of VLASS,
however, the synergies with the SKA precursor surveys can and should be capitalized upon.
For example, the MIGHTEE survey \citep{2017arXiv170901901J} will provide accurate total flux densities and luminosities for extended AGN, whilst also providing a longer baseline in frequency for spectral index measurement that can be used to infer the physical state of AGN and the environments in which they reside.

From the point of view of the dynamic radio sky, VLASS will also provide a unique snapshot of the Universe some 20 years after the FIRST survey. Current large-area radio transient detection surveys, such as the recent Stripe\,82 survey of \citet{2016ApJ...818..105M}, successfully utilized the ``Epoch 0'' provided from FIRST, as well as other historical VLA-based surveys, as a starting point for identification of newly appearing objects from the first new VLA epochs.  

Much of the science of VLASS depends on identification of radio sources in the optical/infrared. Fortunately, many innovative surveys in these wavebands are planned for the next decade.
In the optical, both the Zwicky Transient Facility \citep[ZTF;][]{2014SPIE.9147E..79S} and the Large Synoptic Survey Telescope \citep[LSST;][]{2019ApJ...873..111I} will provide access to information on optical transients, and LSST will also produce
a deep, multiband galaxy survey over the whole southern sky. With a planned launch date in 2022, the {\em Euclid} mission will survey 15,000 square degrees of sky, mostly within the area covered by VLASS, in the optical and near-infrared \citep[e.g.,][]{2018LRR....21....2A}, resulting in sub-arcsecond resolution imaging and grism spectroscopy of a large fraction of the host galaxies of VLASS sources. 
Spectroscopic surveys with the next generation of 
fibre spectrographs \citep[SDSS-V;][]{2017arXiv171103234K}, the Dark Energy Spectroscopic Instrument \citep[DESI;][]{2013arXiv1308.0847L}, the Prime Focus Spectrograph (PFS) on Subaru \citep{2016IAUS..319...55T}, the 4MOST and MOONS spectrographs at the European Southern Observatory \citep{2012SPIE.8446E..0TD,2012SPIE.8446E..0SC}, and the Maunakea Spectroscopic Explorer \citep[MSE;][]{2016arXiv160600043M} will obtain many millions of galaxy spectra, including the hosts of VLASS sources. These can be used not only for redshifts and classifications, but also for studying the stellar populations and metal contents of the host galaxies. By 2030, we expect $\approx 50$\% of VLASS sources to have reliable photometric redshifts
from a combination of space-based infrared and ground-based optical 
photometry, and $\approx 20$\% to have spectroscopic redshifts from large spectroscopic surveys.

\section{Survey Strategy} \label{sec:survstrat}

\subsection{Context}

NVSS has provided
the wider astronomy community with a reliable reference image of
the radio sky at Gigahertz frequencies for the past 20 years. VLASS provides a radio reference image at
2\farcs5 resolution, 20 times higher linear resolution than NVSS (which covers the same sky area), and with three times the sky coverage and two times better linear resolution than FIRST. VLASS also include wideband spectral and polarimetric information at three epochs. 
With a cumulative survey depth of 70\,$\muup$Jy\,beam$^{-1}$, VLASS will be
$\sim$4 times more sensitive than the NVSS to unresolved sources with
spectral indices $\alpha \sim -0.7$, where $S_{\nu}\propto \nu^{\alpha}$).

\subsection{Science Requirements}

The Science Case outlined in Section 2 was distilled into a set of science requirements that the survey should meet. The {\bf all-sky coverage} was driven by the need to detect rare transient events (Section \ref{sec:transsci}), and to provide the largest possible reference survey for multiwavelength studies. The {\bf depth} was driven by the need for the survey to have a significantly better point source sensitivity in the coadded data than previous all-sky radio surveys. The {\bf three epoch strategy} was driven by the need to have each epoch deep enough to detect most classes of radio transient, and the {\bf 32 month cadence} was driven by the need to give slow transients enough time to rise (or fall) between epochs (Section \ref{sec:transsci}). The {\bf angular resolution} was driven by the need to reliably identify optical counterparts to both normal radio sources and transients, and also to reduce beam depolarization when studying the polarimetric properties of extended radio sources (Section \ref{sec:polsci}). In addition, other technical requirements were applied to the flux density scale and polarization calibration in order to ensure accurate spectral and polarization properties.

\subsection{Description}

\noindent \textbf{Area and Depth:}  The goal of the survey is to cover the entire sky visible to the VLA, with a requirement that at least 90\% of that area be observed per epoch. The full area amounts to $\approx 33,885\deg^2$ (all sky north of declination $-40^\circ$), or 82\% of the
celestial sphere. The sensitivity goal is a 1$\sigma$ rms depth of 70\,$\muup$Jy for the three epochs combined. Recognizing that the rms is a function of location on the sky (in particular, in regions with a large amount of interference, the sensitivity may be worse), the survey requirement is that the rms should be within 40\% of the theoretical rms for at least 90\% of the area observed.

\begin{figure*}
    \centering
    \includegraphics[scale=1.0]{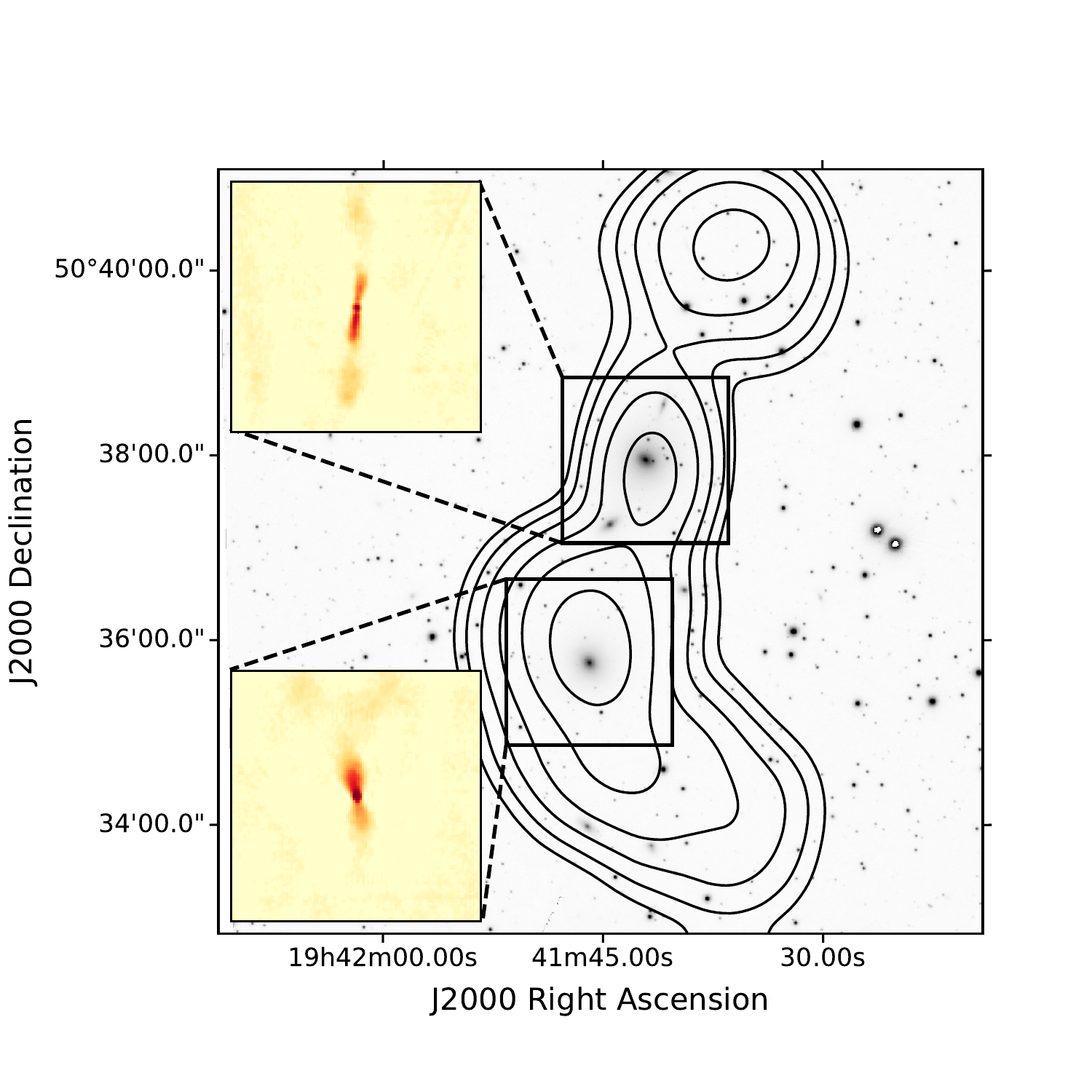}
    \caption{
    The radio source 3C402, illustrating the advance in angular resolution provided by VLASS over the whole sky. The contours show the NVSS image (contour levels 1,2,4,8,16 and 32 mJy), which is superposed on a greyscale of the PanSTARRS $i$-band image. The two
    insets show the VLASS data (made with the lower 1/3 of the frequency band (peak flux density$=$0.062 Jy\,beam$^{-1}$, beamsize $2\farcs9 \times 2\farcs6$ at a position angle of 19$^\circ$) corresponding to the two AGN that make up this radio source \citep{1975MmRAS..80..105R}. The inset boxes are 1.8 arcmin on a side. The resolution of VLASS shows that what looks plausibly like a single radio source in the NVSS breaks up into two separate sources in VLASS, identified with two different galaxies, both at a redshift of 0.026. (Note that 3C402 is outside of the FIRST footprint.)}
    \label{fig:3C402}
\end{figure*}

\vspace{6pt}
\noindent \textbf{Angular Resolution:} High angular resolution is a key requirement for VLASS. It is essential for providing physical insight into the nature of the radio source identifications (for example, whether a transient source is centered on the nucleus of a galaxy or in its periphery, or whether a double source is associated with a single host galaxy nucleus or is a dual AGN), and to correctly identify radio sources either in crowded cluster fields
(e.g., Figure \ref{fig:3C402}), or when the radio source is associated with a very high redshift/dusty AGN that may be almost invisible even in a very deep optical survey. In this regard, \cite{2015ApJ...801...26H} make a case for high angular resolution to make reliable cross-identifications in the next generation of deep optical/near-infrared surveys. The methodology of Helfand et al.\ has, however, been disputed by \cite{2015arXiv150205616C}. Also, \cite{2012MNRAS.423..132M} studied the effect on the reliability of source identifications of decreasing resolution of a radio survey from 6 to 15 arcsec, showing only a modest decrease in completeness of source identifications, and only a small fractional
contamination by incorrect identifications (from 0.8\% at 6-arcsec resolution to 2.3\% at 15 arcsec). Nevertheless, this small fraction of mis-identifications may represent some of the most scientifically interesting objects.

Our angular resolution requirement is therefore driven by the need for VLASS to be able to decompose radio structures on the scale of the optical size of host galaxy at cosmological distances (redshifts $\ga 1$), i.e., $\approx 30\,$kpc, corresponding to $\approx 3$-arcsec. Thus, VLASS is being conducted in the VLA's B and BnA configurations, providing a synthesized beam of $\approx 2\farcs5$ over the entire sky. In practice, the beam varies with declination and hour angle(s) of an observation, leading to the survey requirement that the geometric mean of the FWHM of the synthesized beam should be $<$\,3 arcsec over at least 60\% of the survey, with a beam axial ratio (major axis over minor axis) of $<$\,2.

\vspace{6pt}
\noindent \textbf{Cadence of Multiple Epochs:} A cadence between epochs of at least 2 years is required to meet the transient science goals. Since the VLA configuration cycle is 16 months, observing the same half the sky observable by the VLA each time the VLA is in B-configuration gives a cadence of 32 months for any given point on the sky ($\pm 4$ months to allow for the duration of each B-configuration), and matches well both the scientific and operational needs. The entire sky will be imaged three times down to an rms depth of $\sim$120\,$\muup$Jy/beam on this timescale, providing three epochs of high-resolution maps and catalogs for immediate time-domain science, as well as providing a critical baseline for all future transient surveys and follow-up of multi-wavelength transient events (e.g., gravitational waves, LSST, etc.). 

Observing campaigns are named by epoch and sky coverage, for example, VLASS1.1 observations correspond to the first epoch observations of the first half of the sky, and VLASS1.2 the first epoch observations of the second half of the sky. Thus the final set of observations planned correspond to VLASS3.2.

\vspace{6pt}
\noindent \textbf{Sensitivity:} 
The sensitivity of a single epoch is set by the desire to image the whole sky as rapidly as possible given the data rate limit for regular VLA observations of $\sim 25$\,MB/s. The sensitivity of the VLA for
mosaicking is computed using the procedure given in the VLA mosaicking guide: https://go.nrao.edu/vlass-mosaicking.
For continuum (Stokes I) at 2--4~GHz (S-band) these calculations yield a survey speed, $SS$, of:
\begin{equation}
SS = 16.55 \left( {\sigma_I \over 100\,\muup\textrm{Jy/beam}} \right)^{2}\, \textrm{deg}^2\,\textrm{hr}^{-1}
\end{equation}
for an image rms of $\sigma_I$.
This assumes 1500~MHz of useable bandwidth (after RFI excision) and
an image averaged over the band using multi-frequency synthesis.
(The integration time needed to survey a given area to a particular depth is given by dividing that area by the survey speed.) The data rate constraint implies a maximum survey 
speed of 23.83 square degrees per hour, which results in an RMS of 120$\,\muup$Jy/beam per epoch, or 70$\,\muup$Jy/beam when all three epochs are combined. This is significantly deeper than any previous all sky radio survey. 

\vspace{6pt}
\noindent \textbf{Calibration:} the flux density scale calibration accuracy requirement
is 10\% (with a goal of 5\%) over 90\% of the observed area. 
The polarization leakage requirement is $<0.75$\% (with a goal of 0.25\%) over 90\% of the survey. The absolute polarization position angle accuracy requirement was originally set to 2$^{\circ}$. This was obtained by the need to measure a typical rotation measure of 10 rad\,m$^{-2}$ to 30\% accuracy by comparing a VLASS measurement to a hypothetical perfect measurement at another frequency. Unfortunately, systematic issues with calibrator models and polarization position angle calibration mean that these requirements may need to be relaxed for at least some fraction of the final survey products (see Section \ref{sec:polcal}).

\vspace{6pt}
\noindent \textbf{Overhead:} In the estimates for total observing time,
allowance is made for the overhead for slewing, set-up, and calibration
that will apply to a given component of the survey. When originally planning 
the survey, it was assumed that
the overheads would be $\approx 19$\%, arising from a combination of
4-hr, 6-hr, and 8-hr scheduling blocks. Accounting for additional integration time needed to overcome the increase
in system temperature at low elevations, and a potential failure
rate of 3\% (mostly due to high winds, system failures, or interrupts by Target of Opportunity Observations (ToOs)), a wall-clock time of 5520 hours was predicted to be needed
to execute the survey. This amounts to 920 hours per configuration
cycle. The properties of VLASS based on these assumptions are summarized in Table~\ref{tab:all-sky}. The actual overhead realised during the observations of the first half of the first epoch was 21\%, corresponding to 935 hours per configuration cycle. Calibration overhead was slightly higher, about 23\% for the second half of the first epoch due to the preference for shorter scheduling blocks during the windy season, and the replacement of 3C~48 as a calibrator due to its flaring behaviour at that time.

\section{Survey Implementation}  \label{sec:impl}

The implementation plan is split into an observing plan, a plan for data processing and data products, and archiving and data distribution. More details on the survey implementation may be found in the VLASS Memo Series, available at https://go.nrao.edu/vlass-memos. In addition, we refer later in the text to several of the EVLA Memos available at https://go.nrao.edu/evla-memos.

\subsection{Observing}\label{sec:observ}

\subsubsection{Observing Status}
\label{sec:obssched}

\begin{figure*}
\begin{center}
\includegraphics[angle=-90,scale=0.7,trim=4cm 3cm 4cm 3cm, clip=True]{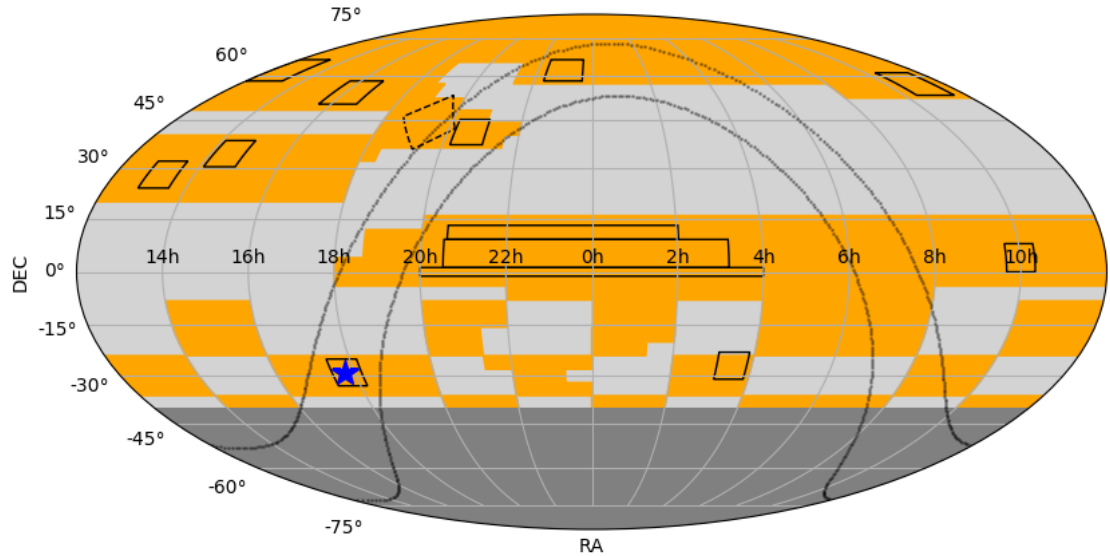}
\caption{\footnotesize Sky coverage for the first and second half of each VLASS epoch.  Orange-shaded areas indicate the regions to be observed in the first half of each VLASS epoch. The light gray areas will be observed in the second half of each epoch. Solid black outlines show the VLASS Pilot Survey fields. The dashed-outline region indicates the Kepler field. The Galactic center is indicated by the blue star; the Galactic plane ($\pm 10^{\circ}$ galactic latitude) is indicated by the dotted curves. The dark grey below Declination $-$40\degr\ is outside the VLASS coverage area.
\label{fig:coverage}
}
\end{center}
\end{figure*}

VLASS as proposed requires approximately 5520 hours of observing time,
which will be carried out over the course of six cycles
of VLA observing in its B and BnA configurations, spanning a total period of
seven years (although the execution of the third epoch is subject to the successful passing of a review during the second epoch). In addition, a $\sim$200~hr pilot survey was undertaken in the summer of 2016 to test the observation strategy, as detailed in VLASS Memo \#2. Observations of the survey proper started in September 2017 as the VLA entered its B-configuration. Throughout the lifetime of VLASS, the VLA will 
continue to follow its 16-month standard configuration cycle (D-C-B-A), with the time in each configuration modified to minimize the impact of the VLASS on other users of the VLA. South of declination $-15^{\circ}$, VLASS is utilizing the hybrid BnA configuration (where antennas on the north arm of the VLA are located in their A-configuration positions, while the east and west arms remain in the B-configuration) between the B and A parts of the standard VLA configuration cycle. This provides a more circular synthesized beam, comparable to that obtained at higher declinations in the B configuration.

Half of the sky was observed in the first configuration cycle, from September 2017 to February 2018 as shown in Figure~\ref{fig:coverage}, and the remaining half of the sky from February 2019 to July 2019, to complete the first all-sky observations. These two halves (campaigns VLASS1.1 and VLASS1.2) together comprise the entire first epoch of VLASS. To maintain the required $\sim$32-month cadence, the parts of the sky that constitute the first and second halves will continue to be observed as the first and second half of every epoch of VLASS. 

\subsubsection{Scheduling Considerations}

The VLA is normally operated using a ``dynamic scheduling queue'' where the individual Scheduling Blocks (SBs) are created in the VLA Observation Preparation Tool (OPT) for a prescribed range of Local Sidereal Time (LST), and submitted to the VLA Observation Scheduling Tool (OST) to be queued up for observation by the array according to weather and scheduling priority. 
To facilitate this for VLASS, the sky is divided into 899 tiles roughly $10^{\circ}\times 4^{\circ}$ in size, each corresponding to an observing duration of $\approx 2$\,hr. To reduce calibration overheads, these tiles are combined (in groups of 2, 3, or 4 tiles) to make scheduling blocks of 4--8\,hr duration. The tiles are arranged in 32 tiers in Declination, from South to North, and are indexed from one in order of increasing RA. The tile names give the tier and the tile number within that tier. For example, the tile named T14t05 is the 5th tile in the 14th tier. See VLASS Memos \#4 and \#7 for details.

Provision is made to accommodate the possibility that time critical,
triggered, target of opportunity (ToO) programs may interrupt a
VLASS observation, by observing all essential calibrations at the start of an SB if at all possible;  there is currently no mechanical provision in
the way schedules are constructed or executed for the suspension
and restarting of schedule blocks on the VLA. The possibility of SB interrupts is minimized by having the majority of VLASS SBs no more than $\sim$4\,hr in length. With this block length, ToO observations
can generally be fitted into the regular dynamic schedule between
VLASS blocks. No VLASS SBs had to be interrupted to support ToOs during VLASS Epoch 1 observations.

\subsubsection{Mosaic Observing Pattern}

In order to carry out \hbox{VLASS}, the sky is observed using a large number of mosaicked pointings of the VLA\@. At 2--4~GHz, the VLA has a field of view given by the primary beam response of the 25-meter diameter antennas (EVLA Memo \#195). In order to optimally cover a given sky area in an efficient manner, the array conducts raster scans using OTFM. In OTFM there is very little move-and-settle overhead as the array is in continuous motion over a row of a raster with only a short startup ($\sim$5--10 sec) at the start of a row. 
The separation of scan rows is chosen by the desire to obtain uniformity in sensitivity. The VLASS row separation of 7.2$^{\prime}$ results in a uniformity within 0.28\% at 3 GHz and 5.1\% at 3.95 GHz. To reach the survey speed goal of 23.83 square degrees per hour (excluding calibration overheads), a scan speed of $\approx$3.31 arcmin$\,{\rm s^{-1}}$ is used. During OTFM, the phase center is discretely sampled often enough (more than ten times across the beam) to ensure minimal ($< 1$\%) smearing of the response. In the case of VLASS, the phase center of the array is discretely stepped every 0.9 seconds, i.e., every 0.3$^{\prime}$. Short correlator ``dump'' times are required to sample these phase centers (0.45 seconds for VLASS). The scan rate is decreased at very low declinations (by 5\% at $\delta=-20^{\circ}$ to 74\% at $\delta=-40^{\circ}$) in order to increase the time-on-sky at low declinations to account for the increased system temperatures when pointing close to the horizon due to ground pickup.

\subsection{Data processing and products}\label{sec:datatech}

VLASS data products are broken down into classes of ``Basic'', ``Enhanced'', and ``Commensal''. The Basic Data Products (BDPs) for VLASS are those produced by \hbox{NRAO} using standard data processing systems or modest extensions thereof. The BDPs have been defined to be a set of products from which the community can derive the science from the survey, requiring little post-processing on the user's part. Enhanced Data Products (EDPs), and Enhanced Data Services (EDSs), require extra resources or involve specialized domain expertise, and are defined and produced by community members. Commensal data products are produced by dedicated backend instruments that operate alongside ongoing VLASS observations (see Section~\ref{sec:commensaldata}).

In order to keep up with the observing, the VLASS BDP Pipeline must process and image the data at a rate commensurate with the observing rate. This is effected through the parallel production of $1^\circ \times1^\circ$ images on NRAO-based cluster nodes or externally provided systems \citep[e.g., through XSEDE;][]{6866038}.

\subsubsection{Basic Data Products} \label{sec:basicdata}

The BDPs of VLASS consist of
\begin{itemize}
\item Raw visibility data;
\item Calibration products and a process to generate calibrated data;
\item Quick Look continuum images;
\item Single Epoch images, and image cubes in Stokes I,Q,U, coarsely sampled in frequency;
\item Single Epoch component catalogs;
\item Cumulative VLASS images, and image cubes in Stokes I,Q,U with both coarse and fine sampling in frequency for epoch 2 and following, and
\item Cumulative VLASS component catalogs.
\end{itemize}
The resources for processing, curating, and serving the BDPs are provided
by \hbox{NRAO}. 
Teams led by \hbox{NRAO} carry out the activities required for the processing and Quality Assurance (QA) of the products.
Table~\ref{tab:bdp} summarizes the components of each data product and
the timescale for their production.  Specific details of individual
BDP are described below.

\begin{table*}
\begin{center}
\caption{VLASS Basic Data Products\label{tab:bdp}}
{\footnotesize
\begin{tabular}{lll}
\noalign{\hrule\hrule}
\textbf{Data Product} & \textbf{Components} & \textbf{Production Timescale} \\
\noalign{\hrule}
Raw Visibility Data & Standard VLA data & Immediate \\
\noalign{\hrule}
\multirow{5}{0.15\textwidth}{Calibrated Data} & Initial Calibration and Flagging Tables & \multirow{5}{0.15\textwidth}{1~week after observation (initial), 3~months after observation (final)} \\
        & Final Calibration and Flagging Tables & \\
        & Pipeline Control Scripts & \\
        & Flagging Commands   & \\
        & QA Reports \& Plots & \\
\noalign{\hrule}
\multirow{2}{0.15\textwidth}{Quick Look Images} & Stokes~I Images & \multirow{2}{0.15\textwidth}{2~weeks after observation} \\
        & Stokes I Noise Images & \\
\noalign{\hrule}
\multirow{4}{0.15\textwidth}{Single Epoch Images} & Stokes I continuum images, \& noise images &
 \multirow{4}{0.15\textwidth}{6~mos.\ I wideband continuum, 12~months\ IQU cubes$^*$} \\
        & Stokes IQU coarse (128~MHz) spectral cubes, \& noise cubes & \\
        & Stokes I Spectral Index, and uncertainty images & \\
        & \quad (generated using Multi Frequency Synthesis) & \\
\noalign{\hrule}
\multirow{7}{0.15\textwidth}{Single Epoch Component Catalog} & Component Position and uncertainty (centroid of Stokes~I emission) &
 \multirow{7}{0.15\textwidth}{with Single Epoch Images} \\
        & Peak Brightness in Stokes~\hbox{IQU} \& uncertainty & \\
        & Flux Density in Stokes~\hbox{IQU} \& uncertainty & \\
        & Spectral Index at Peak Brightness (I) \& uncertainty & \\
        & Integrated Spectral Index (I) \& uncertainty & \\
        & Peak IQU spectrum (bright sources) & \\
        & Basic Shape information for I & \\
\noalign{\hrule}
\multirow{5}{0.15\textwidth}{Cumulative VLASS Images} & Stokes I continuum images, \& noise images &
 \multirow{5}{0.15\textwidth}{12~months\ I wideband continuum, 16~months\ IQU cubes} \\
        & Tapered Stokes I continuum images, \& noise images \\
        & Stokes IQU coarse (128~MHz) spectral cubes, \& noise cubes & \\
        & Stokes IQU fine (16~MHz) spectral cubes, \& noise cubes & \\
        & Stokes I Spectral Index, Curvature, and & \\
        & \quad uncertainty images & \\
\noalign{\hrule}
\multirow{8}{0.15\textwidth}{Cumulative VLASS Component Catalog} & Component Position and uncertainty (centroid of Stokes~I emission) &
 \multirow{8}{0.15\textwidth}{with Cumulative VLASS Images} \\
        & Peak Brightness in Stokes~\hbox{IQU} \& uncertainty & \\
        & Flux Density in Stokes~\hbox{IQU} \&  uncertainty & \\
        & Spectral Index, Curvature at Peak Brightness (Stokes~I) & \\
        & \quad and uncertainty & \\
        & Peak IQU spectrum (bright sources) & \\
        & Basic Shape information for I & \\
\noalign{\hrule\hrule}
\end{tabular}
}
\hbox{$^*$The Single Epoch products will be delayed for epoch 1.}
\end{center}
\end{table*}

\vskip 6pt
\noindent
\textbf{Raw Visibility Data:}
The raw visibility data for VLASS are stored in the standard VLA archive. These data are available immediately after observation, with no proprietary period.

\vskip 6pt
\noindent
\textbf{Calibration Data:}
the VLASS Calibration determines, on the basis of \textit{a priori} factors and from observations of standard calibration sources, the corrections to the raw data amplitude, phase, and visibility weights to be applied to the data. This process also determines the flags that are needed to remove bad data due to instrumental faults, RFI, and other causes of error. When applied to the VLASS data, this calibration allows the production of images in the next processing stage. This process only includes the derivation of the complex gain and bandpass calibration factors known through previous measurements or determined by the observations of calibrators and transferred to the VLASS target observations. Details
of the calibrations applied may be found in Section \ref{sec:earlycal}.

VLASS data are processed using a modified version of the standard CASA-based VLA calibration pipeline, initially developed and tested on the VLASS pilot observations. Once the pipeline has run, and after QA, users can gain access to the calibrated data from the Archive. Archiving the calibrated visibility data themselves would double the required data storage, which is too costly. Instead, the calibration products are archived as BDPs, and the archive maintains scripts to regenerate the calibrated dataset via a request through the NRAO archive interface. 

\vskip 6pt
\noindent
\textbf{Quick Look Images:}
the identification of transient and variable objects is a key VLASS
science goal. Certain populations of transients will evolve on short time scales ($\sim$days) and can only be robustly identified through rapid follow-up VLA observations soon after detection in VLASS data. These include Galactic transients (e.g., stellar flares) and relativistic transients viewed on-axis or close to on-axis (e.g., GRBs, blazar flares). For this reason, Quick Look images are produced as soon as possible after data acquisition, with the goal that they be produced within two weeks of observation. The Quick Look Stokes~I images, averaged over the entire band, and a corresponding rms sensitivity (noise) image are produced using the standard CASA mosaicking algorithm. They are then
made available via the web either from a webserver or the NRAO archive. The simple imaging algorithms used to generate the Quick Look images on a short timescale 
result in relatively poor positional and flux density accuracies when compared to pointed observations of the same fields. The 
properties of the Quick Look images are described in VLASS Memo \#13.

\vskip 6pt
\noindent
\textbf{Single Epoch Images:}
these fully calibrated and quality-assured images will be produced
using a specialized CASA-based wide-field Imaging Pipeline developed by NRAO staff. Data products will comprise Stokes I continuum images (averaged over the band), images of the rms noise, spectral index and uncertainty, and spectral image cubes with 128\,MHz channels for Stokes IQU\@. The planes of these cubes will be stored with their default (native resolution) beams
for their respective frequencies; scripts will be provided to convolve these images to the lowest common resolution ($\approx$3$\farcs5$). 
Currently, details of the processes and algorithms to generate the Single Epoch images are still being finalized. They will be detailed in an upcoming VLASS Memo.

\begin{figure*}[!t]
\begin{center}
\includegraphics[width=18cm]{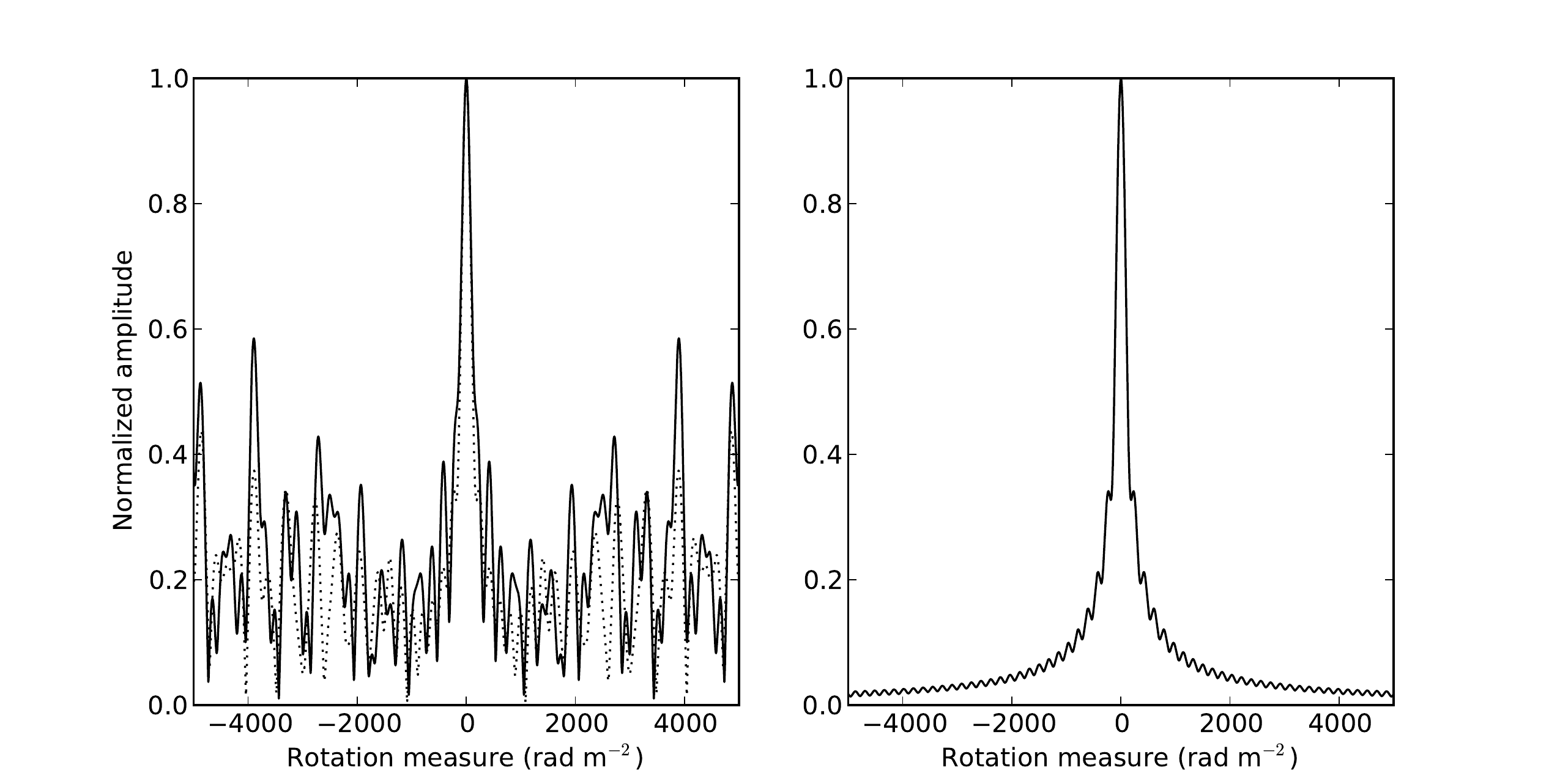}
\caption{\footnotesize
The Faraday depth response function. The dotted line in the left-hand plot shows the response for equally-weighted channels in the VLASS BDP ``coarse'' cubes (128~MHz channel width), the solid line shows the response weighted by the channel noise in a typical Observation from the VLASS Pilot Survey in Stripe 82. The right-hand plot shows the response for the VLASS BDP ``fine'' cubes (16MHz channel width), assuming equally weighted channels. In all cases, the available non-RFI contaminated frequencies at 2--4~GHz provides a FWHM response of $\sim$200 rad~m$^{-2}$, but the fine cubes have a much better response to high rotation measures ($> 1000\, {\rm rad\, m^{-2}}$).}
\label{fig:faraday}
\end{center}
\end{figure*}

\vskip 6pt
\noindent
\textbf{Single Epoch Component Catalogs:}
the main image analysis task for VLASS is the production of the component catalogs for the Single Epoch images. Studies of the performance of different radio continuum image source finders include \citet{2012MNRAS.422.1812H} and \citet{2015PASA...32...37H}, which consider the available options in the context of ASKAP\@. Also available as a proof of concept is the source finding carried out for the VLA Stripe-82 surveys by \citet{2016ApJ...818..105M}. There is also a comprehensive discussion by \citet{2013ApJ...768..165M} on the analysis of archival VLA ECDFS multi-epoch data. We tested several different source detection and characterization software packages using early VLASS data. We found that pyBDSF \citep{2015ascl.soft02007M} provided the best results overall for the VLASS dataset, in particular, 
its ability to reject false detections from sidelobes was found to be superior to that of other codes.
Inclusion of the spectral index images and polarimetric images from VLASS requires some extensions to the basic source finder. More advanced catalogs, including multi-wavelength matching, will be made as part of the EDPs. Details of the Single Epoch component catalogs and tests of different source finding algorithms will be published in a future VLASS Memo.

\vskip 6pt
\noindent
\textbf{Cumulative VLASS Images:}
the Cumulative Images will be produced using similar algorithms to those of the Single Epoch Images, but using all available epochs; these will serve as the final set of VLASS images. The products will comprise Stokes I continuum images (averaged over the band), images of the rms noise, spectral index, spectral curvature and associated uncertainty maps, and spectral image cubes with 128~MHz channels for Stokes IQU, as for Single Epoch. In
addition, there will be finely-sampled (16~MHz channel) cubes for the brightest sources, providing a significantly better Faraday transfer function than the Single Epoch cubes (Figure \ref{fig:faraday}). Also, a set of wideband Stokes I continuum, spectral index, spectral curvature and associated uncertainty maps will be derived using a {\em uv}-taper to improve the surface brightness sensitivity for extended emission (resolution $\approx 7\farcs5$). (Spatial scales up to $\sim 20''$ are well-sampled by the VLA at 3~GHz in the B-configuration for the observing mode used by the survey.) Strongly
variable sources will be taken into account during the imaging for all Cumulative data products. 

\vskip 6pt
\noindent
\textbf{Cumulative VLASS Component Catalog:}
the Cumulative Component Catalog will be produced in a manner similar to the Single Epoch Component Catalog, and will serve as the reference catalog for \hbox{VLASS}.

\subsubsection{Enhanced Data Products}
\label{sec:enhancedata}

The EDPs require additional resource
and/or more domain expertise, and so are defined and produced by the
VLASS community outside NRAO\@. These data products require external
support to define, produce, and validate. 

\smallskip
\noindent
{\bf The CIRADA Project.} Many of the EDPs and EDSs will be provided as part of the Canadian Initiative for Radio Astronomy Data Analysis (CIRADA; http://cirada.org) a
program led by a consortium of Canadian universities, in partnership with NRAO and the Canadian Astronomy Data Centre (CADC), and funded by the Canada Foundation for Innovation. CIRADA will produce EDPs for VLASS as follows:

\begin{itemize}

\item Multi-wavelength catalogues that provide optical spectroscopy/redshifts and
morphology, infrared photometry and X-ray luminosities for each detected
radio object based on available wide-area, multi-wavelength surveys;

\item Identification of clustered detected components into single physically associated radio sources;

\item Average level of Faraday rotation measure for each detected
source;

\item Decomposition of interfering Faraday components in the same source
using rotation measure synthesis;

\item Cubes of Faraday depth across extended sources;

\item Automated quality assurance for BDP images, to identify and correct
common image artifacts that could otherwise act as false positives in
searches for radio transients;

\item Source extraction and catalog generation, featuring rapid and robust identification of variable and transient objects;

\item The VLASS Transient Marshal (VTM), a web interface incorporating
existing multi-wavelength data, allowing users to filter and identify
transients relevant to their own interests;

\item Transient alerts via VOEvent (http://bit.ly/voevent)
which will be automatically transmitted to the wider astronomical community
as soon as possible after completion of the associated VLASS observations, including
a link to the corresponding VTM entry.

\end{itemize}

The CIRADA EDPs and EDSs will use software operating 
on computing infrastructure provided by Compute Canada.
The core areas of work will be:
\begin{itemize}

\item Advanced continuum software development. This EDP captures all
features of the imaging data, including multi-wavelength properties and
spectral slopes of sources across the VLASS band. 

\item Transient Source Identification Software Development: CIRADA will
develop software that automatically assesses data quality and that
identifies time-variable sources on time scales from milliseconds to years.
This EDP also requires the creation of an alert service and event list for
all variable and transient objects.

\item Faraday Rotation and Rotation Measure Synthesis: CIRADA will make Faraday rotation and rotation measure synthesis maps of all polarized sources detected in VLASS.

\item Cross-Matched Database: CIRADA will develop a project-wide database
structure that supports high-performance, distributed queries.  The database
will integrate data from other wavebands (e.g., optical, X-ray) and allow
for cross-matching of EDPs with other astronomy surveys. CIRADA will adapt
existing remote visualization solutions \citep[such as CARTA, ][]{2015ASPC..495..121R} to provide responsive, in-browser visualization databases with $>10^7$ rows.

\item Advanced Analytics, Data Science and Visualisation: CIRADA will build
custom interfaces to the cross-matched database, supporting queries based on
source type using a Bayesian classification framework. This will also
require data mining and visualization tools that exploit the EDPs. 

\item Long-term Archiving and Services: CIRADA will provide enhanced access to
and long-term archiving of science-ready data products. The Canadian Astronomy Data Centre (CADC) is already serving the VLASS Quick Look images via its standard interface, including an image cutout service (https://www.cadc-ccda.hia-iha.nrc-cnrc.gc.ca/en/search/).

\end{itemize}

\noindent
{\bf Other initiatives.} The South African Inter-University Institute for Data Intensive Astronomy \citep[IDIA (http://idia.ac.za);][]{7530650} will play a strong role in the production of polarization products and visualization capabilities, in collaboration with CIRADA.

Other areas for EDPs will undoubtedly become apparent. Criteria for including EDPs in the VLASS archive will be relevance to the VLASS science case and cost of curating and serving the products. 

\subsection{Archiving and Data Distribution}
\label{sec:archivedata}

A comprehensive survey like VLASS produces a diverse set of data products and requires a full-featured archive to serve it to the public. Currently, products are served from a website hosted by the NRAO. This will be replaced in late 2019 by an archive site that will feature basic search capabilities for all the BDPs (raw data, calibration products, catalogs and image products). 

EDPs will be made available through the NRAO archive as resources permit. The VLASS products, either in basic form, or further processed, may also be made available via alternative EDSs (see Section \ref{sec:enhancedata}). EDSs are required for the curation and distribution of EDPs that fall beyond the capabilities of NRAO to support.

\section{Commensal Experiments}
\label{sec:commensaldata}

Independently of VLASS, two groups are developing new instruments dedicated to analyzing VLA data in parallel with all ongoing observations (called ``commensal'' observing). The \textit{realfast} \citep{2018ApJS..236....8L} and VLITE \citep[][]{2016SPIE.9906E..5BC} commensal projects will enhance the scientific productivity of VLASS by performing fast transient searches and low-frequency observing, respectively.

\subsection{Realfast} 
{\em realfast} is a commensal fast transient survey system at the VLA. The VLA correlator has observing modes that generate fast-sampled visibilities at a rate on the order of one TeraByte per hour. However, this data rate is far too large to sustainably record and analyze. \textit{realfast} solves this problem by using a dedicated compute cluster for real-time detection of candidate transients.

As a commensal system, primary observers like the VLASS team configure their observations and retrieve their data from the archive as usual. The VLA observing system will configure the correlator to send a fast-sampled copy of visibilities (sampled every 5-20~ms) to the \textit{realfast} cluster, which dedisperses and images to search for transients over the full field of view \citep{
2017ascl.soft10002L}. The VLA image $1\sigma$\ sensitivity is 5~mJy in 5~ms for observations from 1 to 5~GHz with 1~GHz of bandwidth.

\textit{realfast} started development in late 2016 with support from the National Science Foundation and will begin science observing in the summer of 2019. The initial deployment used existing CPUs in the VLA correlator, but the full system uses a new, dedicated GPU cluster. Candidate transient detections made during VLASS will be integrated with transient alert and other VLASS data aggregation services. The raw visibility data, calibration products, and search software will be openly available to the community (http://realfast.io).

\subsection{The VLASS Commensal Sky Survey}
The VLITE system uses the VLA low band receivers located near the prime focus to observe simultaneously with all regular Cassegrain observations. Developed with funding from the US Naval Research Laboratory (NRL), VLITE uses splitters, dedicated samplers and spare fibers to make available 64~MHz of bandwidth covering  320--384~MHz from the new broadband low frequency receivers \citep{2011ursi.confE...5C} to a custom-designed real-time DiFX correlator producing full polarization products \citep{2007PASP..119..318D}. It operates at a spectral resolution of 100~kHz and temporal resolution of 2\,s.

VLITE became operational in November of 2014.  During its initial 2.5 years of operation, VLITE recorded more than 15,800 hours of data using 10 antennas.  In 2016, VLITE's DiFX correlator was modified to allow data recording during OTFM to provide support to VLASS\@. In 2017, VLITE was expanded to 16 antennas, which more than doubles the number of baselines. VLITE also has an independent fast transient GPU-based processing system which has recently begun operation. This system is capable of detecting fast radio bursts if the dispersion measure is less than 1000~pc~cm$^{−3}$.

Standard VLITE data are processed for astrophysics use and transient searches with a two day delay using an automated pipeline based on the Obit data reduction package \citep{2008PASP..120..439C}. Pipeline astrophysical data products include science-ready images, calibrated uv data and source catalogs.  The VLASS data are processed separately to allow survey-specific processing choices, such as common restoring beams.  The main VLITE data products are three degree square science-ready mosaics; however calibrated uv-data and source catalogs are also kept.

\begin{figure*}
\centering
\includegraphics[scale=0.6,trim=4cm 0cm 0cm 0cm,clip=True,angle=-90]{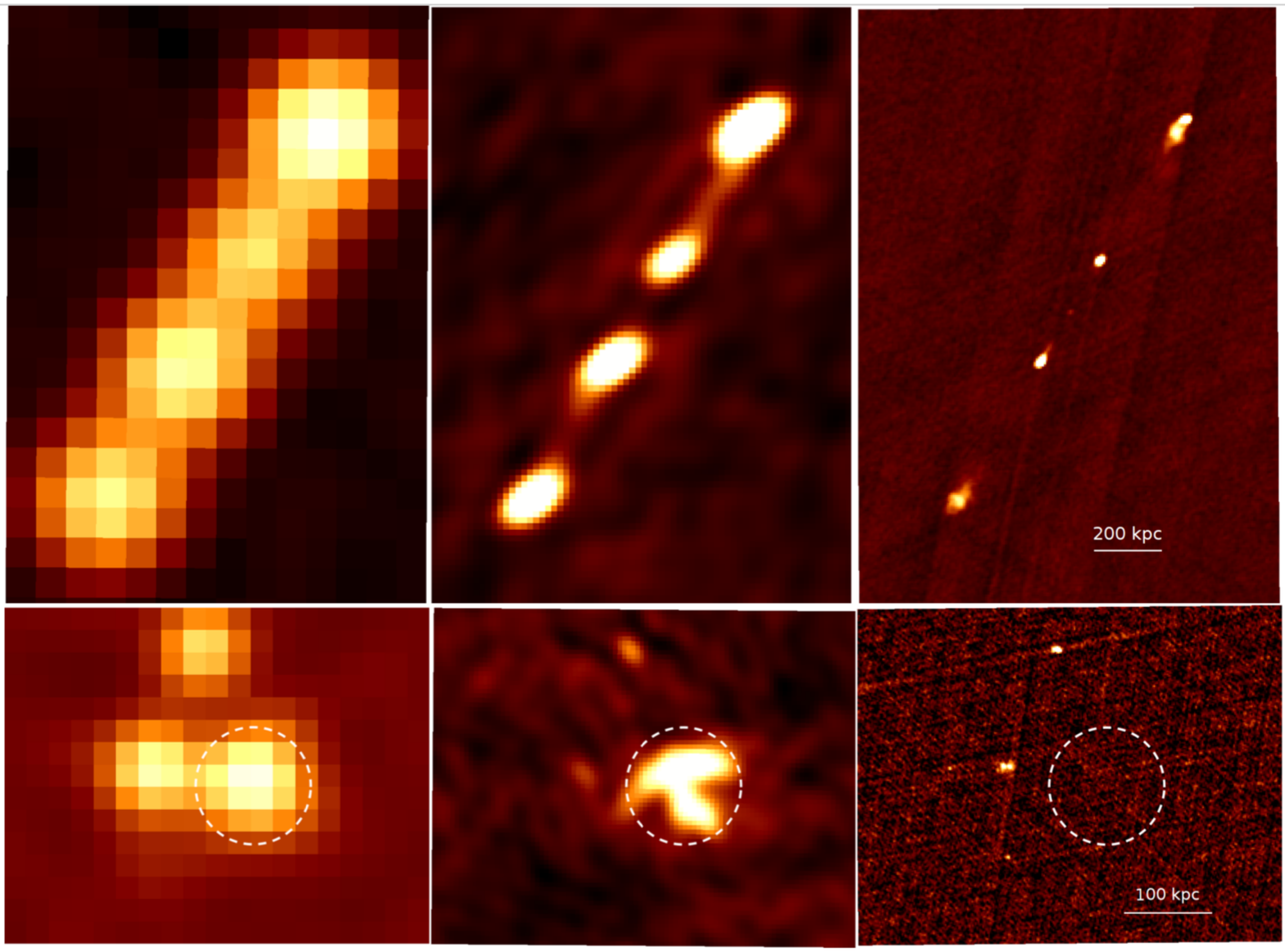}
\caption{VCSS will add the detection of low surface brightness steep
spectrum emisson to the VLASS dataset. The top row shows the $z=0.519$ double-double radio galaxy B1834+620 in NVSS (left, peak flux density = 240\,mJy\,beam$^{-1}$), the VCSS epoch 1 first look image (middle, peak flux density = 660\,mJy\,beam$^{-1}$, beam=23$\farcs$13$\times$13$\farcs$6, at a position angle of $-58^\circ$) and the VLASS first epoch Quick Look image (right, peak flux density = 38\,mJy/beam, beam=2$\farcs$88$\times$2$\farcs$13, at a position angle of $-59^\circ$). The white scale bar on VLASS image shows 200~kpc at the redshift of the source. The bottom row shows the field near the center of the galaxy cluster Abell 566 ($z=0.097$) in NVSS (left, peak flux density=20\,mJy/beam), VCSS epoch 1 (middle, peak flux density = 122\,mJy/beam, beam=18$\farcs$32$\times$12$\farcs$20, at a position angle of $26^\circ$), and the VLASS first epoch Quick Look image (right, peak flux density = 4.3\,mJy/beam, beam=2$\farcs$74$\times$2$\farcs$14, at a position angle of $40^\circ$). The white scale bar on the VLASS image shows 100 kpc at the redshift of the galaxy cluster. The diffuse halo source (circled) is detected in NVSS and VCSS, but not in VLASS.
\label{fig:VCSS}}
\end{figure*}

VLITE was operational for the VLASS pilot observations during the summer of 2016 using ten antennas, and initial processing of the data resulted in mosaics with noise levels of $\sim$6\,mJy/beam with a resolution of $\sim$20~arcseconds. During the first epoch of survey observations in 2017, VLITE operated with 15 to 16 antennas. Automated data processing is underway, and initial mosaics have noise levels at roughly 3\,mJy/beam and resolutions of $\sim$12--25~arcseconds. Observations from the VLITE for VLASS Commensal Sky Survey (VCSS), when combined with the VLASS data, offer instantaneous spectra for point sources across the sky, can serve as an independent verification of any transient events, and provide information on extended source structures. We highlight the science potential of VCSS in Figure~\ref{fig:VCSS} where we show two examples of sources seen in NVSS, VCSS and VLASS. The top row highlights the resolution of VCSS over NVSS for separating components of a double-double radio galaxy, enabling a more direct spectral comparison to VLASS. In the bottom row, VCSS reveals the structure of a steep spectrum source in a galaxy cluster that is unresolved in NVSS and not detected by VLASS. This shows that VCSS will be of great value for detecting diffuse objects (with scale sizes $\stackrel{>}{_{\sim}}30$ arcsec) that will be resolved out by VLASS. VCSS is also more sensitive to ultra-steep spectrum ($\alpha < -1.8$) sources than VLASS.

NRL archives all raw and calibrated VLITE data and provides it to the community for scientific use (details at https://go.nrao.edu/vlite). VCSS data products may eventually be served from the NRAO archive. The continued expansion of the VLITE system into a full 27 antenna wideband commensal system, called the LOw Band Observatory (LOBO), may occur during the VLASS program.

\section{Calibration of VLASS data}\label{sec:earlycal}

\subsection{Overview}
Calibration of VLASS uses a modified version of the CASA calibration pipeline that is designed for ALMA and VLA data \citep{2015ASPC..499..355S}.
The VLASS calibration pipeline is now fairly mature, and unlikely to change significantly over the course
of the project, so we describe it here. Imaging of VLASS data is still under development. Quick Look images are now available for the pilot and first epoch, and are characterized in VLASS Memo 13.\footnote{Draft available at {\url  https://go.nrao.edu/vlass-memo013}} The Single Epoch and Cumulative imaging pipelines are yet to be finalized, and will be detailed in future VLASS publications. 

The steps in the calibration pipeline follow standard practice in radio interferometry. To begin, the data are flagged for known issues caught by on-line systems during the observations. Then, the model of the flux density calibrator is obtained. (The primary flux density calibrator used is 3C\,286, though when this is not available other common calibrators such as 3C\,48 or 3C\,138 are used.)  Standard calibrations for antenna gain curves, atmospheric opacity and antenna positions are then made. Any compression of the data due to RFI is corrected for using the switched power information (Section \ref{sec:compression}). First-pass calibration tables are then made for the delay, bandpass and gain calibrations. RFI and bad data on the bandpass and delay calibrators are flagged. The calibrations are then recalculated, and then a further iteration of flagging on the calibrators, this time including the gain calibrators, is performed. The flux densities and spectral indices of all calibrators for which prior models are not available are obtained, and the final 
set of standard calibrations is made and applied. At this point, the polarization calibration (Section \ref{sec:polcal}) is derived, and any outliers in the calibration tables are flagged before they are applied to the data.
One more round of data flagging is then carried out, to flag the target data. Finally, the 
weights of the visibilities are reset according to the standard deviation in each 
spectral window. This often helps with reducing the effects of low level RFI that are missed by the earlier flagging.

\subsection{Gain compression}\label{sec:compression}

RFI can lead to significant gain compression \citep[e.g.,][]{2016ApJ...818..105M}, especially in regions of the sky dominated by satellite emission. This can lead to erroneous flux calibration, and, in particular, some parts of the survey near the ``Clarke Belt'' (corresponding to the apparent Declination ($\approx -5.5^{\circ}$) of geostationary satellites as seen from the VLA) are badly affected by this. We therefore implemented a correction that uses the switched power data (EVLA Memo \#145). 
The switched power adds a known power $P_{\rm cal}$ for half the time, switched in and out at a rate of 10~Hz. By taking the difference of the power when the switched power is on
($P_{\rm on}$) 
to that when it is off ($P_{\rm off}$), the system gain can be estimated as $G=(P_{\rm on}-P_{\rm off})/P_{\rm cal}$ (assuming the other contributions to the power, including the RFI, remain steady on this timescale). This correction was applied to about 30\% of the scheduling blocks.

This allows calibration of the gain as a function of time, and thus correction of the gain compression (when it affects the data by about 30\% or less). Careful scheduling ensured that gain compression levels of $>$30\% were only seen in $<0.1$\% of the data (see EVLA Memo \#206).

\subsection{Polarization calibration} \label{sec:polcal}

VLASS is observed with full polarization calibration. During VLASS1.1, leakage terms were derived either from observations of a point source over a large range of parallactic angle ($\ga$60~deg), or via observations of a source known to be unpolarized. From VLASS1.2 onwards only observations of an unpolarized calibrator will be used, in order to simplify scheduling and calibration. (Pilot observations were also observed with polarization calibrators where possible, but few observations have a large enough range in parallactic angle to calibrate well.) The polarization position angle is determined using observations of the primary calibrator 3C286 (or, if this is not available, 3C138 or J1800+7828). It should be noted that all the calibration is performed on-axis, and that the quality of the polarization calibration of the images will deteriorate away from the phase centers. The level of deterioration will be assessed prior to the generation of Single Epoch polarization products, and may result in a decision being made to correct it using polarization-dependent primary beams.

Commissioning of VLASS polarization calibration revealed an unexpected limitation in determining accurate polarization position angles, described in EVLA Memo \#205. A residual variation in circular polarization cross-hand phases that depends on parallactic angle is observed, which translates to an increased uncertainty of the polarization position angles. Unfortunately, this does not seem to be correctable. The absolute polarization position angles thus cannot be assumed reliable to better than $\sim$5$^\circ$. This does not affect the accuracy of relative polarization position angles, and is thus not expected to significantly degrade the ability to determine rotation measures within the VLASS data itself, however, it will affect comparisons with other datasets.

The data products are produced in Stokes parameters I, Q and U. Although circular polarization (Stokes V) is, in principle, also calibrated, the calibration systematics in Stokes V are much larger and it is not considered usable for science. 
The width of the Faraday transfer function (when the full 2--4\,GHz bandwidth is available) is 200~rad\,m$^{-2}$ (Figure~\ref{fig:faraday}), and the uncertainty in the rotation measure is $\approx 200/(2\cdot\mathrm{SNR})$~rad\,m$^{-2}$. The uncertainty level was confirmed by measuring the standard deviation of fitted rotation measures to bright point sources with simple Faraday screens in the Stripe 82 data from the pilot survey.

\subsection{Calibration Performance}
\label{sec:qa}

\subsubsection{The flux density scale}

\begin{figure*}
    \centering
    \includegraphics[scale=0.4,angle=0]{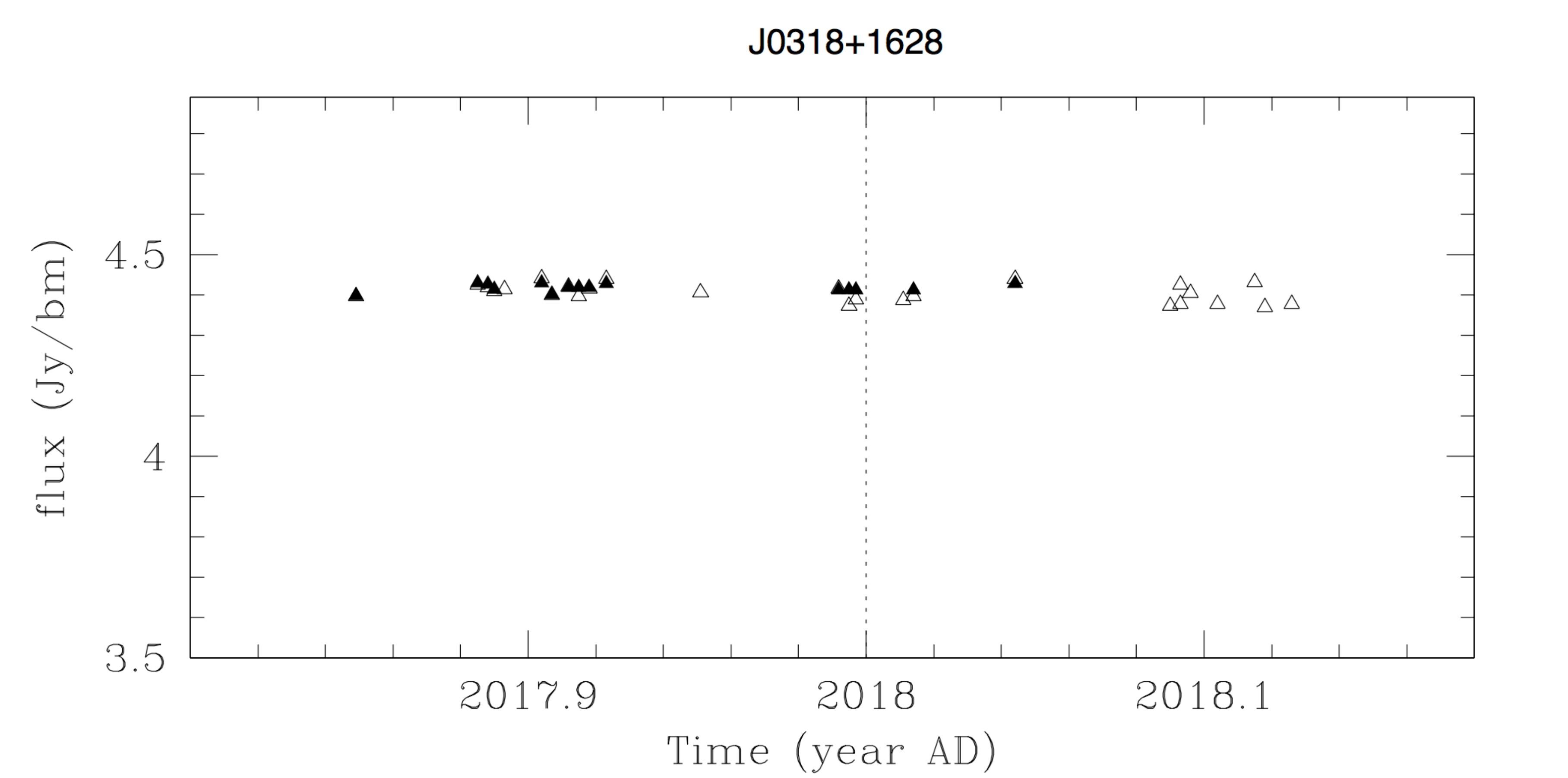}
    \includegraphics[scale=0.4,angle=0]{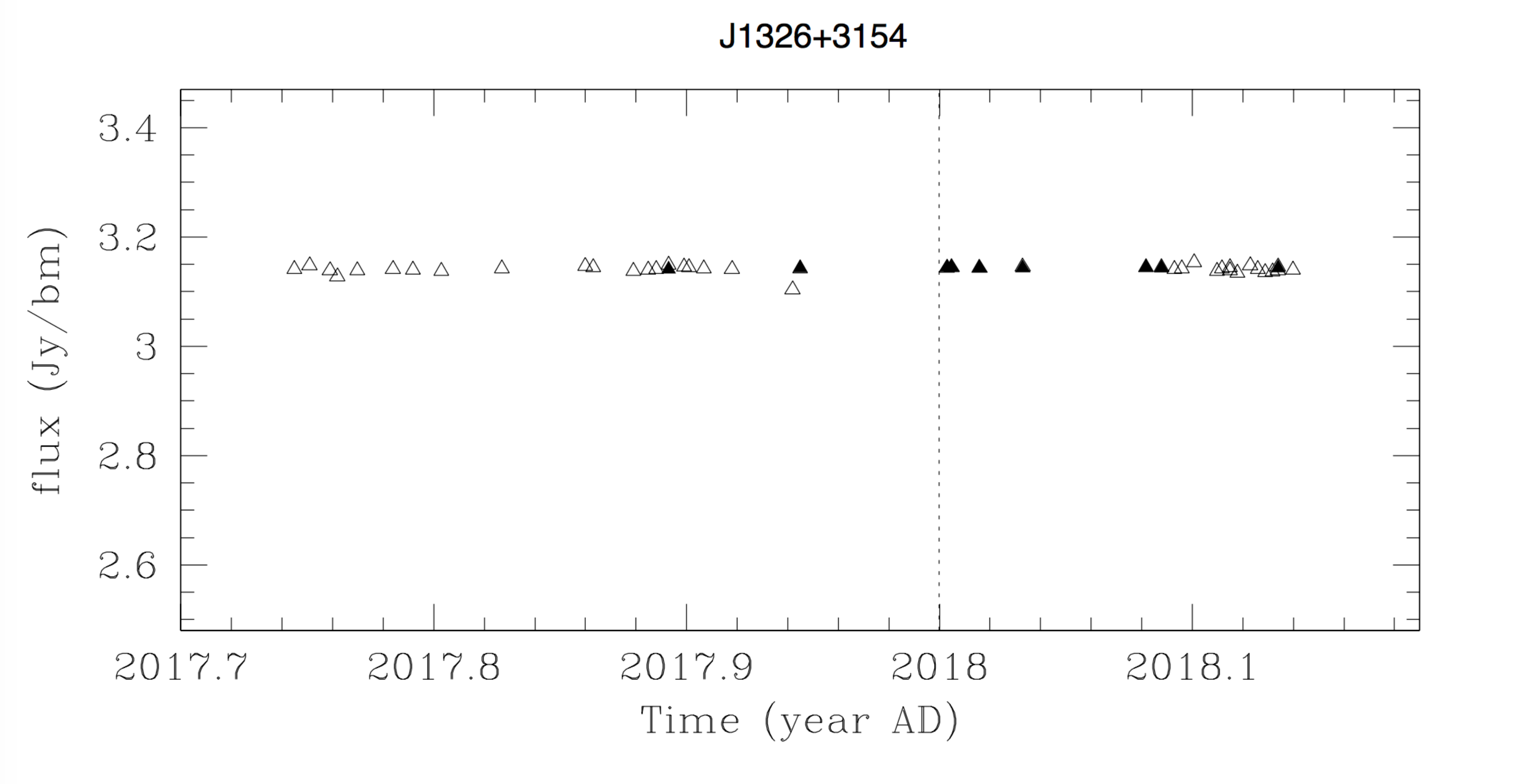}
    \caption{Two phase calibrators, J0318+1628 and J1326+3154, that showed little variability when observed multiple times during the first half of the first epoch. Open triangles represent observations that did not need to be corrected for gain compression, filled triangles correspond to observations that had the correction applied.}
    \label{fig:calibrators}
\end{figure*}

Several phase calibrators were observed multiple times during VLASS 1.1, as they were used to calibrate more than one tile. These can be used to give an estimate of the reproducability of the flux density calibration (if the phase calibrator is not variable). Two of these that appeared not to be variable, J0318+1628 and J1326+3154, were selected to check the stability of the flux calibration, and the accuracy of the correction
for gain compression. The RMS about the mean is 1.9\% for J0318+1628 (26 observations) and 0.7\% for J1326+3154 (42 observations), indicating that the flux calibration consistency is very good (Figure \ref{fig:calibrators}). Note, however, that, as discussed in VLASS Memo \#13, the Quick Look images do not achieve this level of flux density accuracy due to the imaging algorithm used.

\begin{figure*}
\includegraphics[width=13cm,angle=-90]{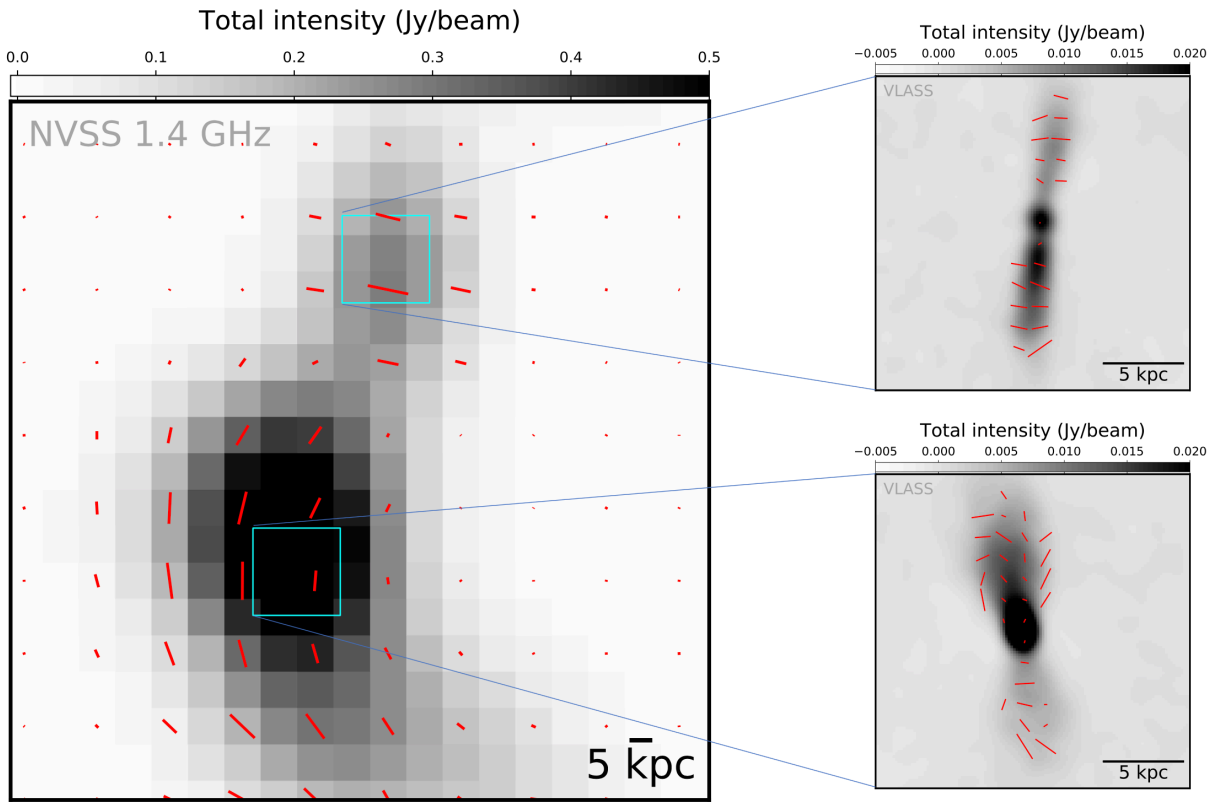}
\caption{
Polarization vectors superposed on total intensity images of the radio source 3C402. On the left is the NVSS data, on the right are the two active radio
galaxies 3C402N and 3C402S that together make up the NVSS source. The NVSS total intensity image peaks at 0.71\,Jy/beam. In the NVSS image, the lengths of the polarization vectors are scaled such that a 5\,kpc long vector (the size of the scale bar) corresponds to a fractional polarization of 1\% in the low-band (1365~MHz) polarization image of \cite{2009ApJ...702.1230T}. In the VLASS images, 5\,kpc  corresponds to a fractional polarization of 86\%.
The beam of the NVSS image is $64\farcs \times 54\farcs$ at position angle 1$^{\circ}$; the beam of the VLASS image is $2\farcs9 \times 2\farcs6$ at position angle 19$^{\circ}$.}
\label{fig:3C402_polvec}
\end{figure*}

\begin{figure*}
\includegraphics[width=13.0cm,angle=-90]{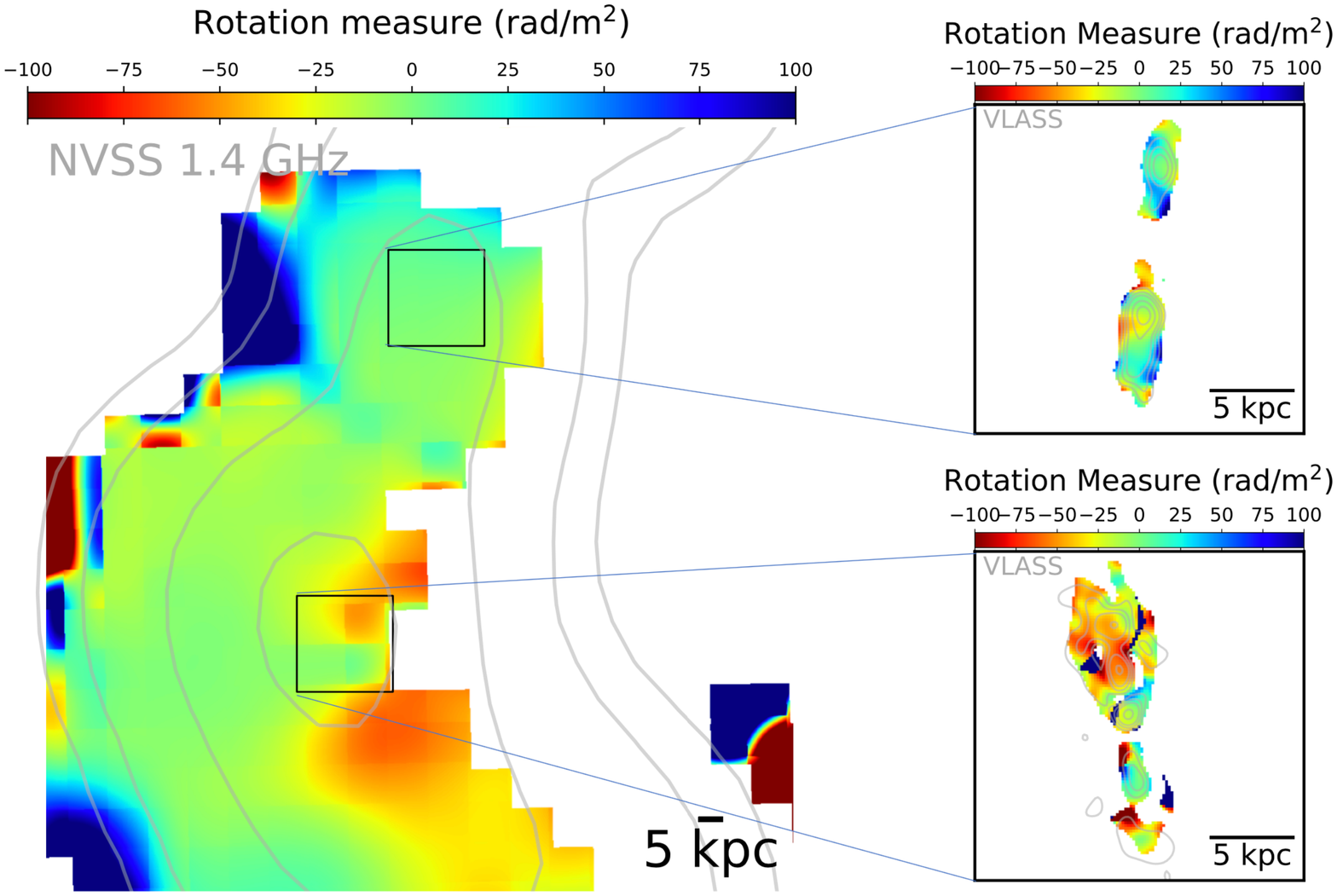}
\caption{Rotation measure images for 3C402. On the left is the NVSS image, using reprocessed data from \citet{2009ApJ...702.1230T}. Contours are
Stokes I flux density (levels 9, 30, 150, 450 and 840 mJy/beam). On the right are shown 3C402N and 3C402S in VLASS, with contours of polarized flux density (levels 0.8, 1.2, 1.6, 2.4, 3.2 Jy/beam).}
\label{fig:3C402_rm}
\end{figure*}

\subsubsection{Polarization}\label{sec:obspol}

Early polarimetry data from the first epoch of VLASS are shown in Figures \ref{fig:3C402_polvec} and \ref{fig:3C402_rm}. These show VLASS polarimetric data on 3C402 (also shown in total intensity in Figure \ref{fig:3C402}) compared to split-band data from NVSS \citep{2009ApJ...702.1230T}. The much higher angular resolution of VLASS enables variations in the Faraday Rotation to be seen that were not apparent in the NVSS data. Overall, the rotation measures (RM) from NVSS and VLASS agree well. For the Northern component (3C402N), we obtain $-1.8$~rad$\,$m$^{-2}$ from NVSS and $-1.6\;$rad$\;$m$^{-2}$ (with a scatter of $30\,$rad$\,$m$^{-2}$) from VLASS (variations in the RM in the NVSS image are not meaningful as the entire source is contained within one beam). For the southern component (3C402S) we obtain $-28$~rad$\;$m$^{-2}$ from NVSS and $-33\,$rad$\,$m$^{-2}$ (with a scatter of $25\,$rad$\,$m$^{-2}$) from VLASS. As expected, given the much reduced beam depolarization, the fractional polarization is much higher in the 
VLASS images. 3C402N is 6.7\% polarized in NVSS, compared to 13\% (with a scatter of 8\%) from VLASS, and for 3C402S is 1.7\% polarized in NVSS compared to 11\% (with a scatter of 8\%) from VLASS. 

\section{Education and Outreach} \label{sec:EPO}

Education, outreach and communication have been included in the planning for VLASS from the outset and are an integral part of the survey. Information for scientists and the public about VLASS and access to available material are disseminated via multiple platforms (social and print media, presentations, etc.) and include links to the VLASS web page at vlass.org and to NRAO Science web pages.  These provide access to the catalogues and data products for astronomers, whether expert radio astronomers or not. 

Formal and informal education and public outreach (EPO) activities, e.g., blogs, images and community events, are listed at vlass.org.  Our emphasis is on the development of programs that are evidence-based, i.e., are firmly rooted in education research and evaluation methods, and involve both astronomers and educators in all phases of the project.  Specific programs may be new or leverage already existing ones reflecting the interests of  scientists and educators. Therefore we encourage all scientists who use VLASS data and are interested in engaging with the public to contact epo@vlass.org. Examples of the types of activities that will become relevant after the first epoch of data collection are citizen science research, engaging with non-professional science societies such as local astronomy clubs and teacher training programs, and participating in lecture circuits.

As an example of one of our VLASS EPO activities, in the first year of VLASS operations we held two VLASS data visualization workshops. The goal of these activities was for the participants to learn how to produce images from radio observations that balance aesthetic considerations with scientific information. Participants learned techniques for constructing images that engage the viewer such that they turn to resources to learn more about the astronomy targets.  
While the resultant images are intended to explain the science, the
{\em process} of creating public images is exploratory, and can lead to scientific discoveries. Many journal examples show that the clarity of the image is appealing for professional articles as well. 
  
We describe the first workshop, held 2018 July 11, since it introduces to radio astronomers two departures from usual practices.  This was the first time the instructor for this workshop was remote and instructing via an online video application (Zoom). The 20 participants, consisting of summer interns, postdocs, faculty and staff, had low through medium imaging skill levels. Nevertheless the experiment was a success, in part due to the assistance of a moderator, who monitored the chat dialogs and progress. 

\begin{figure*}
\centering
\includegraphics[scale=0.5]{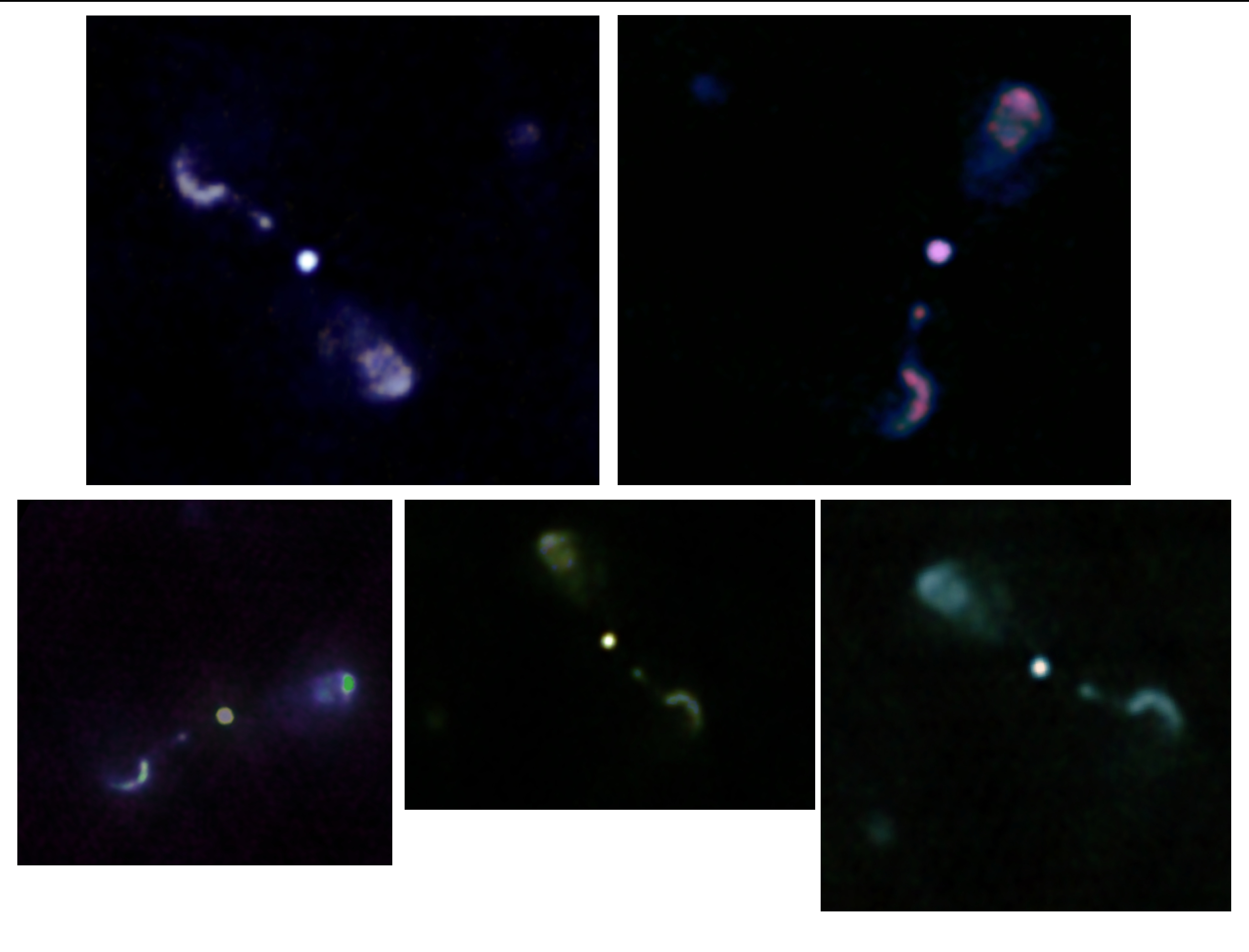}
\caption{\small 
Images produced during an image-making workshop at the University of Manitoba (July 2018). Using the layers schema in GIMP, a different colour was assigned to each of four VLASS frequency subbands for the AGN 4C~48.49. Each participant attempted to retain the viewers' attention by selecting a harmonious palette of colours, orientation, and cropping.  A mask was applied to reduce Quick Look imaging artifacts.  The variety of images demonstrates that a number of visualization solutions are valid. Credit: Top row, left to right:  G. Ferrand (RIKEN) and D. Romano (University of New South Wales). Bottom row, left to right: M. Boyce, Y. Gordon, A. Vantyghem (University of Manitoba).
\label{fig:UMVLASSwkshp}}
\end{figure*}

A preparatory lecture introduced the participants to the methods of visualization, human perception, visual grammar, color harmonies and conventions expected in public outreach images. (Information  on image-making techniques are provided in \citet{2017IJMPD..2630010E}.) 
Subsequently, participants worked collaboratively in small teams for enhanced learning via social engagement. They used Quick Look VLASS images of the radio source 4C~48.49 divided into 4 subbands, and the free GNU Image Manipulation Package (GIMP). Rather than assigning different colors to a single parameter, they assigned a color to each of the subbands. This second departure from a convention in radio astronomy, allowed participants to reveal the changes in spectral index across the target without using contour-like plots. During the critique portion of the workshop the participants considered whether the color harmony they chose communicated the science they wished to highlight to a layperson.  Although each team used the same data, the resultant draft images were vivid and diverse.  Those from a second VLASS imaging workshop are shown in Figure \ref{fig:UMVLASSwkshp}. 

\begin{figure*}
\centering
\includegraphics[scale=0.15]{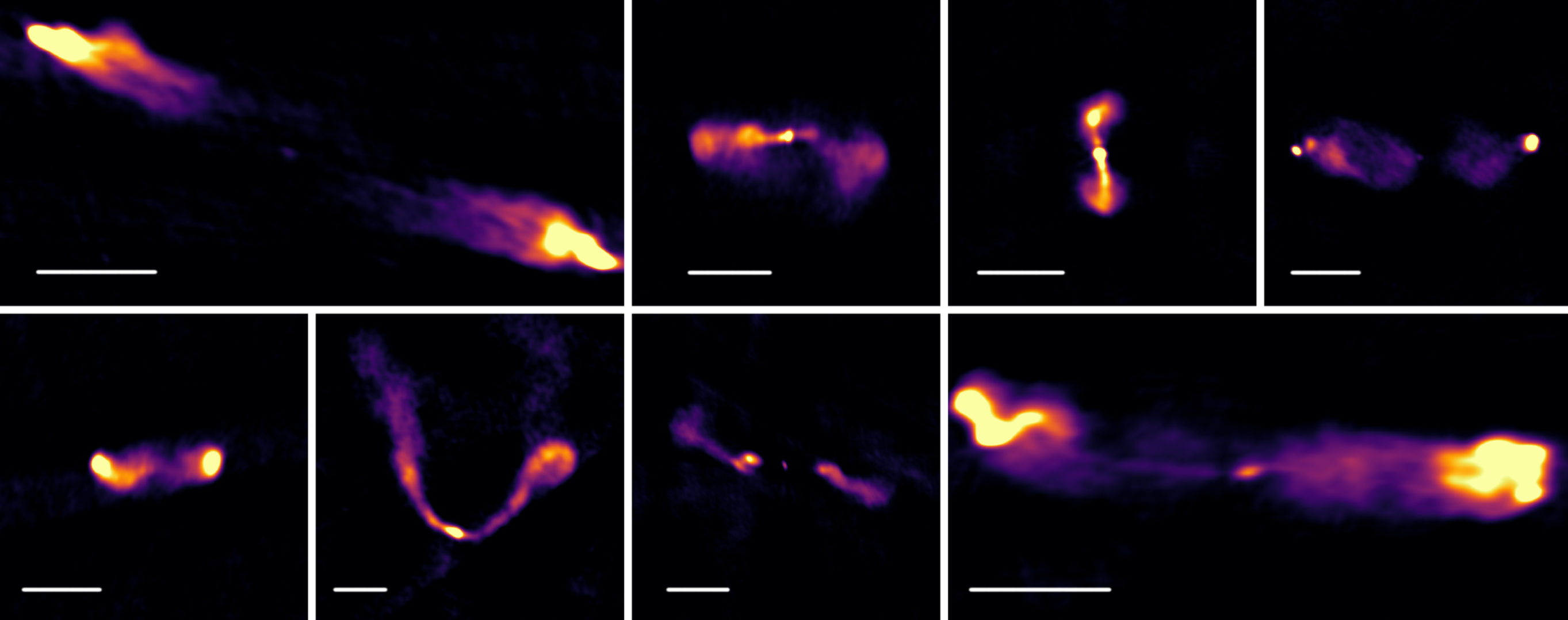}
\caption{\small 
A sample of cutout images of extended VLASS sources identified during a recent NINE workshop by participants from the University of the West Indies, St.\ Augustine. The scale bars are 30 arcseconds in length.}
\label{fig:NINE}
\end{figure*}

VLASS imaging data and products are also used in Science, Technology, Engineering and Math (STEM) programs for students.  The NRAO National and International Non-traditional Exchange program (NINE) aims to broaden participation of under-represented groups in radio astronomy.  VLASS imaging data products are used as a means to expose students to Python and data mining.   Participants in NINE workshops bring the computing and development skills back to their home institutions to lead tutorials with fellow students and faculty (https://go.nrao.edu/nine). Figure~\ref{fig:NINE} shows the results from a NINE workshop at the University of the West Indies, St.\ Augustine.

\section{Summary} \label{sec:summary}
The VLA Sky Survey is a radio survey with a unique combination of high (2\farcs5) resolution, sensitivity (a goal of 70$\,\muup$Jy rms), full linear Stokes polarimetry, time domain coverage, and wide bandwidth (between 2 and 4\,GHz) over 82\% of the sky. It is the first large-scale synoptic radio survey ever undertaken, and will enable 
a science program that will benefit the entire astronomical community, open new
parts of scientific discovery space, and keep its legacy value for decades to come. It is a fitting complement to large surveys in the optical and infrared domains that are either currently underway, or planned for the next decade.

A pilot survey was undertaken in the summer of 2016, the first half of the first epoch was observed from 2017 September to 2018 February, and the second half of the first epoch from 2019 February to 2019 July. These observations successfully demonstrated the feasibility of the survey, and indicate that it will ultimately reach the goals described in this paper. Quick Look image products are now available for these early observations, and, although their use for science should be subject to strong caveats regarding the positional and flux density accuracy, they have already proven to be useful for some scientific applications. A full suite of Single Epoch processing will commence early in 2020, which will result in significantly more accurate data products and include polarimetry.

\acknowledgements

The National Radio Astronomy Observatory is a
facility of the National Science Foundation operated under cooperative
agreement by Associated Universities, Inc. CIRADA is funded by a grant
from the Canada Foundation for Innovation 2017 Innovation Fund (Project 35999), 
as well as by the Provinces of Ontario, British Columbia, Alberta, Manitoba and
Quebec. This research made use of NASA's Astrophysics Data System (ADS)
Abstract Service. Part of this research was
carried out at the Jet Propulsion Laboratory, California Institute of
Technology, under a contract with the National Aeronautics and Space
Administration. SC acknowledges support from the NSF (AAG 1815242). 
SC and TJWL are members of the NANOGrav Physics Frontier Center, which is supported by the NSF (award number 1430284. CJL is supported by NSF award 1611606. Partial support for the work of L.\ Rudnick comes 
from NSF grants AST-1211595 and 1714205 to the University of Minnesota.
This work used the Extreme Science and Engineering Discovery Environment (XSEDE), which is supported by National Science Foundation grant number ACI-1548562.
HA benefited from grant CIIC 218/2019 of the University of Guanajuato, Mexico. 
BRK acknowledges support from the NRAO NINE program and Office of Diversity and Inclusion.
GRS acknowledges support from NSERC Discovery Grant RGPIN-06569-2016.
SBS is supported by NSF award \#1458952. SPO acknowledges financial support from the Deutsche Forschungsgemeinschaft (DFG) under grant BR2026/23.
L.L.\ acknowledges the financial support of DGAPA, UNAM (IN112417), and CONACyT, Mexico.

\bibliography{VLASS_bib_v1.bib}

{\it Facilities:} \facility{VLA}.

\end{document}